\documentclass[twocolumn,epsf,psfig]{revtex4-1}
\usepackage[utf8]{inputenc}
\usepackage{amssymb}
\usepackage{epsfig}
\DeclareMathAlphabet{\pazocal}{OMS}{zplm}{m}{n}            
\usepackage{graphicx}
\usepackage{color}
\usepackage{bm}
\usepackage[normalem]{ulem}     
\usepackage{soul}
\usepackage{amsmath}
\usepackage{braket} 
\usepackage{rotating}
\usepackage{multirow} 
\usepackage{lipsum}
\usepackage{hyperref}

\begin{document}
\title{Revealing hidden magneto-electric multipoles using Compton scattering }         

\author{Sayantika Bhowal} 
\affiliation{Materials Theory, ETH Zurich, Wolfgang-Pauli-Strasse 27, 8093 Zurich, Switzerland}    
\author{Nicola A. Spaldin}
\affiliation{Materials Theory, ETH Zurich, Wolfgang-Pauli-Strasse 27, 8093 Zurich, Switzerland}

\date{\today}

\begin{abstract}
Magneto-electric multipoles, which are odd under both space-inversion $\cal I$ and time-reversal $\cal T$ symmetries, are fundamental in understanding and characterizing magneto-electric materials. However, the detection of these magneto-electric multipoles is often not straightforward as they remain ``hidden" in conventional experiments in part since many magneto-electrics exhibit combined $\cal IT$ symmetry. In the present work, we show that the anti-symmetric Compton profile is a unique signature for all the magneto-electric multipoles, since the asymmetric magnetization density of the magneto-electric multipoles couples to space via spin-orbit coupling, resulting in an anti-symmetric Compton profile. We develop the key physics of the anti-symmetric Compton scattering using symmetry analysis and demonstrate it using explicit first-principles calculations for two well-known representative materials with magneto-electric multipoles, insulating LiNiPO$_4$ and metallic Mn$_2$Au. Our work emphasizes the crucial roles of the orientation of the spin moments, the spin-orbit coupling, and the band structure in generating the anti-symmetric Compton profile in magneto-electric materials.    
\end{abstract}

\maketitle

\section{Introduction}  \label{sec0}
Multipoles provide a convenient basis for describing properties as diverse as electric charge densities and gravitational fields, and are therefore widely used in many areas of physics, such as classical electromagnetism, metamaterials, nuclear and particle physics \cite{LandauLifshitz,Jackson,Stoyle,Raab,Kaelberer}. They are particularly useful in condensed matter systems for characterizing the charge, spin and orbital magnetic moments of electrons in a unified way, and have enabled understanding as well as prediction of various cross-couplings and transport properties  \cite{Spaldin2008,Hayami,Spaldin2013,Watanabe2020}. 

Multipoles can be categorized into four groups based on their behavior (odd or even) under time-reversal ($\cal T$) and space-inversion ($\cal I$) symmetries \cite{Dubovik,Spaldin2008,Hayami}. A substantial body of work in materials physics has focused on the even-parity (even under $\cal I$)
multipoles, involving prediction and observation of {\it hidden} higher-order multipoles, beyond the conventional electric and magnetic dipoles, in the localized electrons of $d$- and $f$-electron systems \cite{Onimaru,Cricchio,Santini}. Recently the search has extended to odd-parity multipoles, motivated in part by  predictions of intriguing properties that may emerge from them \cite{Spaldin2008,Hayami,Spaldin2013,Sumita,Watanabe2020,Watanabe2017,Fu}. The magneto-electric (ME) multipoles, which are the subject of this work, form the lowest order member of the family of multipoles that are odd under both space-inversion and time-reversal.  

The ME multipole
tensor, formally defined as ${\cal M}_{ij} =  \int  r_i \mu_j (\vec r) d^3r$, describes the next order of spatial inhomogeneities in the magnetization density of a material, beyond the magnetic dipole. By definition, the components of the tensor break both $\cal I$ and $\cal T$ symmetries, which is the same condition that allows the linear ME effect, in which an applied electric field induces a magnetization, $M_i = \alpha^{\rm ME}_{ij} E_j$, 
and vice versa. Indeed, the linear ME response of a material can be conveniently discussed in terms of the  
three irreducible (IR) components of the ${\cal M}_{ij}$ tensor, which are the scalar magneto-electric monopole $a$, the vector toroidal moment $\vec t$, and the symmetric traceless quadrupole moment tensor $q_{ij}$ \cite{Spaldin2008,Spaldin2013}.   

Direct experimental detection of these ME multipoles is a challenge in that they often remain ``hidden" to conventional probes and analyses. To our knowledge to date, only resonant x-ray Bragg diffraction, nonreciprocal linear dichroism  (magneto-chiral dichroism), and polarized neutron diffraction have shown signatures of ME multipoles \cite{Staub2009,Kimura,Lovesey,Lovesey2019}.
Recently, an additional possibility, Compton scattering, which measures the electron density as a function of momentum, was proposed as a candidate probe for the toroidal moment based on symmetry arguments  relating a non-zero toroidal moment to an antisymmetric Compton profile \cite{Collins2016}. While first-principles calculations indicated the required non-zero antisymmetric Compton scattering in two toroidal materials, GaFeO$_3$ and LiNiPO$_4$, preliminary experiments, were unable to detect a signal above the noise level \cite{Collins2016,LNPO}. A detailed physical understanding of the process, however, which is crucial for identifying appropriate toroidal materials with stronger antisymmetric Compton responses, is still missing. 

Here, we extend the earlier arguments of Ref.~\onlinecite{Collins2016} to show that, in addition to the toroidal moment, the other ME multipoles, $a$ and $q_{ij}$, manifest in the anti-symmetric Compton profile, vastly increasing the possible phase space of eligible materials. Using symmetry analysis, we extract the one-to-one correspondence between each of the ME multipoles and the direction in the momentum space along which the anti-symmetric Compton profile appears. We then develop the key physics relating the anti-symmetric Compton profile to the ME multipoles, and show that the asymmetry in the magnetization density associated with the ME multipoles is mediated by the spin-orbit coupling (SOC) to the band structure asymmetry required for an anti-symmetric Compton profile. Explicit first-principles calculations based on density functional theory for two representative materials, insulating LiNiPO$_4$ and metallic Mn$_2$Au, then allow us to determine the factors -- band structure, fine details of the spin configuration, and SOC strength -- that determine the strength of the anti-symmetric Compton profile, suggesting guidelines for identifying materials with stronger effects. 

Our work highlights the importance of the  non-trivial duality between real and momentum space that causes the ME multipoles to be spin independent and odd order in $k$ in momentum space. As a result, they can be detected in $k$ space using probes that are not spin-resolved. In addition to the Compton scattering explored here, this observation implicates other momentum-space $k$-resolved but spin-insensitive probes, such as angle-resolved photoemission spectroscopy (ARPES), as suitable for direct observation of magnetoelectric multipoles.

\section{Magneto-electric multipoles and anti-symmetric Compton profile} \label{sec1}

We begin with a brief introduction to the ME multipoles, (for a more detailed discussion see for example Refs. \onlinecite{Spaldin2008,Spaldin2013}) with an emphasis on their manifestation in the anti-symmetric Compton profile. Our focus is on the symmetry relations, as well as the non-trivial duality between the real and the momentum space, both of which are important in understanding the anti-symmetric Compton profile. 

\subsection{Magneto-electric multipoles} \label{MEmultipole}
The ME multipoles, defined as ${\cal M}_{ij} =  \int  r_i \mu_j (\vec r) d^3r$, describe the inhomogeneity in the the magnetization density $\vec \mu (\vec r)$ (including both spin and orbital contributions), to lowest order beyond the usual magnetic dipole moment $ \vec m = \int  \vec \mu(\vec r) d^3r$.
In contrast to the magnetic moment  $\vec m$, which breaks only $\cal T$ symmetry, ${\cal M}_{ij}$
 breaks both $\cal I$ and $\cal T$ symmetries as 
$\vec r$ and $\vec \mu$ break $\cal I$ and $\cal T$ symmetries respectively. As a result,  materials with non-zero ${\cal M}_{ij}$ satisfy the symmetry conditions  required to show the linear ME effect, with a
one-to-one correspondence between the matrix elements ${\cal M}_{ij}$ and the ME tensor elements $\alpha^{\rm ME}_{ij}$. The nine-component tensor ${\cal M}_{ij}$, can be decomposed into three IR components, the ME monopole $a$,  toroidal moment $\vec t$, and ME quadrupole moment $q_{ij}$, such that
\begin{widetext}
 \begin{eqnarray} \nonumber    \label{Mtensor}  
{\cal M}= 
 \left[
{\begin{array}{*{3}c}
   a+\frac{1}{2}(q_{x^2-y^2}- q_{z^2}) & t_z+q_{xy}  & t_y+q_{xz}\\
   -t_z+q_{xy} &  a-\frac{1}{2}(q_{x^2-y^2}- q_{z^2}) & -t_x+q_{yz}\\
   -t_y+q_{xz} & t_x+q_{yz} & a+q_{z^2} \\
  \end{array} }   \right]. \\
    \end{eqnarray} 
\end{widetext}

As seen from matrix (\ref{Mtensor}),  the ME monopole $a$ is a pseudo scalar (tensor of rank zero), and contributes to the diagonal elements of the ME tensor,
\begin{eqnarray}  \label{MEa}
a= \frac{1}{3} {\cal M}_{ii} =  \int  \vec r \cdot \vec \mu (\vec r) d^3r,
\end{eqnarray}
while the toroidal moment vector $\vec t$ is a tensor of rank 1 and  constitutes the anti-symmetric part of the  ${\cal M}_{ij} $ tensor, 
\begin{eqnarray}  \label{MEt}
t_i= \frac{1}{2} \varepsilon_{ijk} {\cal M}_{jk} =  \frac{1}{2} \int  \vec r \times \vec \mu (\vec r) d^3r.
\end{eqnarray}
Finally, the five-component traceless quadrupole magnetic moment tensor, $q_{ij}$, contributes to the symmetric part (both diagonal and off-diagonal elements) of the ME tensor
\begin{eqnarray} \nonumber  \label{MEq}
q_{ij} &=& \frac{1}{2} ( {\cal M}_{ij}+ {\cal M}_{ji} -\frac{2}{3} \delta_{ij} {\cal M}_{kk}) \\
 &=&  \frac{1}{2}   \int  \big(r_i \mu_j + r_j \mu_i - \frac{2}{3} \delta_{ij} \vec r \cdot \vec \mu \big)  d^3r .
\end{eqnarray}
\begin{figure}[t]
\centering
\includegraphics[width=\columnwidth]{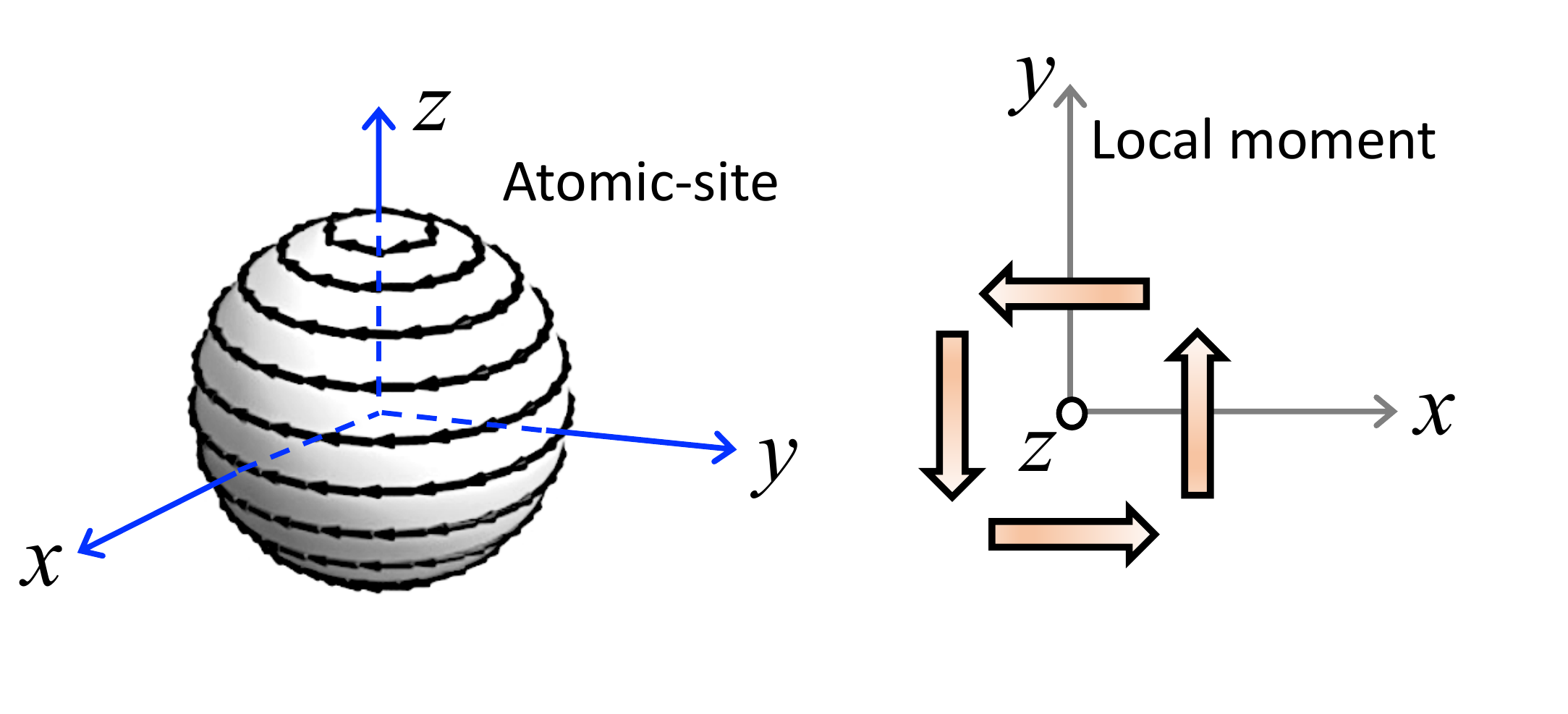}
 \caption{Cartoons showing the atomic-site ($as$) and local moment ({\it loc}) contributions to the ME multipole moments. The illustrations are for the $z$-component of the toroidal moment $t_z$.   The magnetization texture on a sphere around an atom, as shown on the  {\it left}, gives the $as$ contribution to $t_z$. The {\it loc} contribution from the local magnetic dipole moments are shown on the {\it right}.}
 \label{fig0}
 \end{figure}

We point out that each of the ME multipoles in Eqs. (\ref{MEa})-(\ref{MEq}) can be further  decomposed into two contributions, one originating from the magnetization density asymmetry $\vec \mu (\vec r)$ around an atomic site $\alpha$ (known as the atomic site ($as$) contribution) and one from the asymmetric distribution of the local dipole moments $\vec m$ (called the local moment ({\it loc}) contribution) \cite{Spaldin2013}. The two contributions are illustrated schematically for the case of the $t_z$ component of the toroidal moment in Fig. \ref{fig0}. Mathematically, the decomposition for the toroidal moment is given by, \cite{Claude}
\begin{eqnarray} \nonumber \label{as-lm}
	\vec t &=& 2^{-1} \int \vec r \times \vec \mu (\vec r) d^3r \\ \nonumber
	&=& 2^{-1}\sum_\alpha \int_{as}  [(\vec r-\vec r_\alpha)\times  \mu (\vec r) + \vec r_\alpha \times  \vec\mu (\vec r)]  d^3r \\ \nonumber
	&=& 2^{-1}\sum_\alpha \int_{as}  (\vec r-\vec r_\alpha)\times  \mu (\vec r)d^3r + 2^{-1} \sum_\alpha  \vec r_\alpha \times  \vec m_\alpha \\ 
	&=& \vec t_{as} + \vec t_{loc},
\end{eqnarray}
where the summation is over all the atoms that have a local magnetic moment.

\subsection{Anti-symmetric Compton profile} \label{ACP}
Next we show that all the ME multipoles described above contribute to the anti-symmetric Compton profile, and identify the  correspondence between each specific ME multipole and the antisymmetric Compton profile along a particular direction in momentum space.

Compton scattering is an inelastic X-ray scattering process in which the energy loss of the photon is proportional to the projection of the electron momentum density along the direction of the photon momentum transfer. As a result, the measured double differential cross-section $\frac{d^2\sigma}{d\Omega dE}$ along direction $z$ say, is linearly proportional to the Compton scattering profile, $J(p_z)$, which is related to the electron momentum density, $\rho(\vec p)$, by:  
\begin{eqnarray} 
	J (p_z) = \int \rho(\vec p) dp_xdp_y. 
\end{eqnarray}
In materials, where either of the two symmetries, space-inversion and time-reversal, is present or both 
$\rho(\vec p) = \rho(-\vec p)$, and the Compton scattering profile is symmetric in momentum. If both symmetries are broken, however, neither the 
electron momentum density $\rho(\vec p)$ nor the Compton profile $J (p_z)$ are required to be symmetric with respect to $\vec p$. This was  pointed out in Ref. \cite{Collins2016}, where it was suggested that Compton scattering could therefore be used to probe the toroidal moment $\vec t$. 
Here we extend Ref. \cite{Collins2016}, to show that not only the toroidal moments but also all the other ME multipoles can have an antisymmetric contribution to their Compton scattering.

Next we connect the Compton profile to the momentum space representation of the ME multipoles via the multipole expansion of the density matrix.
The parity-odd and time-odd sector of the density matrix is of interest to us, as it corresponds to the 
anti-symmetric part of the Compton profile. 

Therefore, we expand the density matrix $[\rho]$, with respect to its behavior under space inversion and time reversal,
\begin{eqnarray} \nonumber
	[\rho] &=& \sum_{\nu=0,1} \sum_{\eta=0,1} \rho^{\nu\eta}, \\ 
	 {\rm where}~ {\cal T}\rho^{\nu\eta} & = & (-1)^\nu  \rho^{\nu\eta}, ~ {\cal I}\rho^{\nu\eta} =(-1)^\eta  \rho^{\nu\eta}. 
\end{eqnarray}
Expanding the part of the density matrix that is odd under both space and time inversion, $\rho^{11}$, in terms of the ME multipoles in Eqs. \ref{MEa}-\ref{MEq}, we obtain \cite{Spaldin2013}
\begin{eqnarray} 
	\rho^{11} = \sum_{\gamma, \beta}  \Gamma^{11\gamma}_{\beta} T^{\gamma}_{\beta} \quad . 
\end{eqnarray}
Here $\gamma=|\nu-\eta|,..,\nu+\eta = 0, 1, 2$ represents the rank of the parity odd ($\eta=1$) and time odd ($\nu=1$) ME multipoles, discussed in section \ref{MEmultipole}, with $\gamma =0, 1, 2$ corresponding to the ME monopole, toroidal moment, and the quadrupole moment respectively, and the integer $\beta$ runs from $-\gamma,...,\gamma$. Since $\rho^{11}$ describes the component of the density that is odd under both time-reversal and space-inversion, we can conclude that all ME multipoles contribute to the 
anti-symmetric Compton profile. 

Next we analyze the momentum space representations of the time-odd, parity-odd multipoles, in order to establish the directions in which the anti-symmetric Compton profile is non-zero. 
The momentum space representations of the odd-parity multipoles have a different form from their real space representations due to the nontrivial duality between the real space and the 
momentum space. We can see this from the following considerations: The functional form of the spatial part of the odd parity multipoles must be odd order in both $r$ and $k$ space, because both $r$ and $k$ change sign  under inversion  i.e.\  $\vec{r} \overset{\cal I}{\rightarrow} -\vec{r} $, and $\vec{k} \overset{\cal I}{\rightarrow} -\vec{k} $. Under time-reversal, however, the behavior of $r$ and $k$ is different, with $\vec{r} \overset{\cal T}{\rightarrow} \vec{r} $, but $\vec{k} \overset{\cal T}{\rightarrow} -\vec{k} $. As a result, the spin dependence
must be different in the real-space and the momentum space
representations. This leads to the following intriguing consequences for the odd parity multipoles: First,
the basis functions of the odd-parity magnetic
multipoles in momentum space are spin-independent but odd order in $k$. This gives rise to antisymmetric electron (and magnon) dispersions   \cite{Watanabe2017,Watanabe2018} which manifest as the anti-symmetric Compton profile.
Second, the specific basis function of the allowed ME multipole in a system dictates the 
momentum \emph{direction} of the anti-symmetric Compton profile. The detection of the anti-symmetric Compton profile along a certain direction, therefore, can be used to identify the presence of a particular ME multipole in a material.

\section{Anti-symmetric Compton profile in example systems:} \label{sec2}

In order to illustrate the above ideas of the anti-symmetric Compton profile, we explicitly compute and analyze the Compton profile, for two example materials (i) LiNiPO$_4$ and (ii) Mn$_2$Au. While LiNiPO$_4$ is an insulating system and exhibits the linear ME effect \cite{Mercier}, Mn$_2$Au is an anti-ferromagnetic  metal, in which an electric current has been shown to reorient the direction of the magnetization \cite{Shick,Bodnar,Chen,Florian}.  
 The presence of ME multipoles has been demonstrated computationally for both materials, and their relevance for the electric-field or electric-current induced magnetism discussed \cite{Spaldin2013,Florian}. Here, using a combination of symmetry analysis and DFT calculations, we show that the anti-symmetric Compton profile acts as a unique fingerprint of  the symmetry-allowed ME multipoles in these materials.  From a detailed analysis of the DFT results, we extract the role of the magnetic structure, band structure and SOC effects in the anti-symmetric Compton scattering. The insight into the physical mechanism that these provide is a first step to identifying materials with a larger anti-symmetric Compton response.

\subsection{Insulating system: ${\rm LiNiPO_4}$} 

LiNiPO$_4$ crystallizes in the olivine structure with the orthorhombic space group  $Pnma$ and the crystallographic point group $D_{2h}$ \cite{Abrahams}.  The crystal structure, in which the Ni atoms at Wyckoff positions $4c$ are surrounded by six oxygen atoms forming distorted NiO$_6$ octahedra, is shown in Fig. \ref{fig1}(a). 
The material undergoes various magnetic transitions in the presence of a magnetic field, and the resulting phases exhibit linear and quadratic ME effects  \cite{Exp2020}.  Out of these various magnetic phases, the magnetic structure corresponding to the magnetic space group  $Pnm'a$ with the propagation vector $\vec q=(0,0,0)$ constitutes  the magnetic ground state of the system. In this magnetic structure, the Ni atoms are anti-ferromagnetically ordered, with the major spin component along the $z$ direction and a small $x$ component due to canting [see Fig. \ref{fig1} (a)]. The magnetic configuration
 breaks the inversion symmetry of the structure, allowing for ME toroidal and quadrupole moments consistent with  its
 off-diagonal linear ME effect \cite{Jensen,Spaldin2013,Exp2020}. Here, we discuss the manifestation of these allowed ME multipoles in the anti-symmetric Compton profile of LiNiPO$_4$.

\subsubsection{Symmetry analysis}
 \begin{table} [t]
\caption{The basis functions for the ME multipoles for the $D_{2h}$ point group: ME monopole ($a$), toroidal moment ($\vec t$), and the quadrupole moment $q_{ij}$. Here $m_x, m_y, m_z$ are the $x$, $y$ and $z$ components of the magnetic moment $\vec m$, 
that includes both spin and orbital contributions,
$\vec m \equiv \mu_B \big(\frac{2\vec l}{l+1} +2\vec s \big)$.
$\vec l$ and $\vec  s$ are respectively the orbital and the spin angular momentum. }
\setlength{\tabcolsep}{4pt}
\centering
\begin{tabular}{ c c c   c }
\hline
IR    &  ME multipole & \multicolumn{2}{c}{  ~~~~Basis in}     \\
       &                        &   Real space     &  $k$ space \\ [1 ex]
\hline
 $A_u^{-}$          &   $a$,   &   $xm_x+ym_y+zm_z,$  &  $k_xk_yk_z$   \\
                            & $q_{x^2-y^2}$, &  $(xm_x-ym_y)$,  &  \\ 
                &  $q_{z^2}$ & $(2zm_z-xm_x-ym_y)$  &             \\
 $B_{1u}^{-}$     &  $ t_z, q_{xy} $                      &  $(xm_y-ym_x),  (xM_y+yM_x)$   & $k_z$ \\
$ B_{2u}^{-}$     &  $t_y, q_{xz}$                       &   $(zm_x-xm_z),  (zm_x+xm_z)$   &  $k_y$ \\
 $B_{3u}^{-}$    &  $t_x, q_{yz}$                       &    $(ym_z-zm_y),  (ym_z+zm_y)$  &  $k_x$ \\
  \hline
\end{tabular}
\label{tab1}
\end{table}
Before presenting the results of DFT calculations for the anti-symmetric Compton profile in LiNiPO$_4$, we first analyze the symmetry of the magnetic structure and the allowed ME multipoles. This symmetry analysis reveals the direction(s) of the non-zero anti-symmetric Compton profile in momentum space, which we further verify in our DFT calculations presented in section \ref{DFT-LNPO}. 

The real and momentum space basis functions for the time-odd, parity-odd ME multipoles, corresponding to the point group $D_{2h}$, of LiNiPO$_4$, are obtained following Ref. \cite{Watanabe2018} and using the following
compatibility relations ~\cite{books,Bilbao},
the IRs of $D_{4h}$ point group reduce to the $D_{2h}$ point group as
$A_{1u} \downarrow D_{2h} = A_u$, $B_{1u} \downarrow D_{2h} = A_u$, $A_{2u} \downarrow D_{2h} = B_{1u}$, $B_{2u} \downarrow D_{2h} = B_{1u}$, $E_{u} \downarrow D_{2h} = B_{2u} + B_{3u}$. 
and the result is listed in Table \ref{tab1}.

 The magnetic ground state ($Pnm'a$) of LiNiPO$_4$ corresponds to the IR representation $B_{2u}^-$. As seen from Table \ref{tab1}, this $B_{2u}^-$ representation allows for a net toroidal moment along the $y$ direction ($t_y$) and a $q_{xz}$ quadrupole moment. The magnetic point group $mm'm$ at the Ni site allows in addition for the atomic-site ME monopole $a$ and  quadrupole moments $q_{x^2-y^2}$, $q_{z^2}$. These, however, are arranged in an anti-ferro type pattern between the Ni atoms, so that the net values of these ME multipoles are zero. 

We now focus on the basis functions for these ME multipoles, which is crucial for the desired anti-symmetric Compton profile. As expected from the discussion of the real space-momentum space duality in section \ref{ACP},  the real space basis functions in Table \ref{tab1} are spin dependent while the momentum space basis functions are not. More interestingly, the basis functions in the momentum space are always odd order in $k$. For example the momentum space basis for the ME multipoles $t_y, q_{xz}$ (in the $B_{2u}^-$ representation) is $k_y$, indicating an asymmetric dispersion along $k_y$ direction. Thus, for LiNiPO$_4$, we expect an anti-symmetric Compton profile along the $y$-direction in momentum space, as well as along any directions with  $k_y \ne 0$.   

It is interesting to point out here that each of the ME multipoles in Table \ref{tab1} allows for an antisymmetric Compton profile. For example, the existence of the ME multipoles $\{ t_x, q_{yz} \}$ and $\{ t_z , q_{xy} \}$ indicate an antisymmetric profile along $k_x$ (and any direction with $k_x \ne 0$) and $k_z$ (and any direction with  $k_z \ne 0$) respectively. The situation is, however, different for the ME monopole moment $a$ and the  quadrupole moments $q_{x^2-y^2}, q_{z^2}$, which give rise to an antisymmetric profile only along the direction in the momentum space, where  simultaneously $k_x \ne 0, k_y \ne 0, k_z \ne 0$ conditions are satisfied (see Table \ref{tab1}). 
While these symmetry arguments are useful for predicting the {\it directions} in which an anti-symmetric Compton profile will occur, to develop an understanding of the process, and in particular to identify materials with large responses, it is instructive to know the {\it magnitude} of the anti-symmetric profile.

\subsubsection{DFT results for the Compton profile} \label{DFT-LNPO}

In order to verify the above symmetry analysis as well as to gain insight into the physics of the anti-symmetric Compton profile, we next compute the anti-symmetric Compton profile for LiNiPO$_4$ using the linearized augmented plane wave (LAPW) method
as implemented in the ELK  code \cite{code,elk}.  We use the LDA+SOC+$U$ formalism, with $U=5$ eV and $J=0.75$ eV at the Ni site, except in the cases for which we study the effects of varying these parameters.  A basis set of $l_{max(apw)} = 8$, a $3\times6\times6$ k-point sampling of the Brillouin Zone are used to achieve  self-consistency. The product of the muffin-tin radius (2.0, 2.4, 2.2, and 1.8 a.u. for Li, Ni, P, and O respectively) and the maximum reciprocal lattice vector is taken to be 7. All calculations are done at the relaxed atomic positions and lattice constants reported in Ref. \cite{Spaldin2013}. The electron momentum densities 
are calculated and projected onto the selected momentum directions ($\vec p$) to obtain the Compton profile $J(\vec p)$ \cite{elk}. The computed profile is, further, separated into symmetric $J^s (\vec p)$ and anti-symmetric $J^a (\vec p)$ parts, using
$ J (\vec p) = 2^{-1} [J (\vec p)+J (-\vec p)] + 2^{-1} [J (\vec p) - J (-\vec p)]  =J^s (\vec p) + J^a (\vec p)$.
The convergence of the symmetric and the anti-symmetric parts of the profile are confirmed by performing additional calculations with the denser  $5\times10\times10$ k-point mesh.
The computed symmetric and anti-symmetric parts of the profiles are  normalized to the number of valence electrons per formula unit of LiNiPO$_4$ in the calculation, which is 48 electrons. The isotropic core contribution, obtained from the Hartree-Fock calculations
of Biggs {\it et. al.} \cite{Biggs}, is further added to the symmetric part of the profile to obtain the total Compton profile.

\begin{figure}[t]
\centering
\includegraphics[width=\columnwidth]{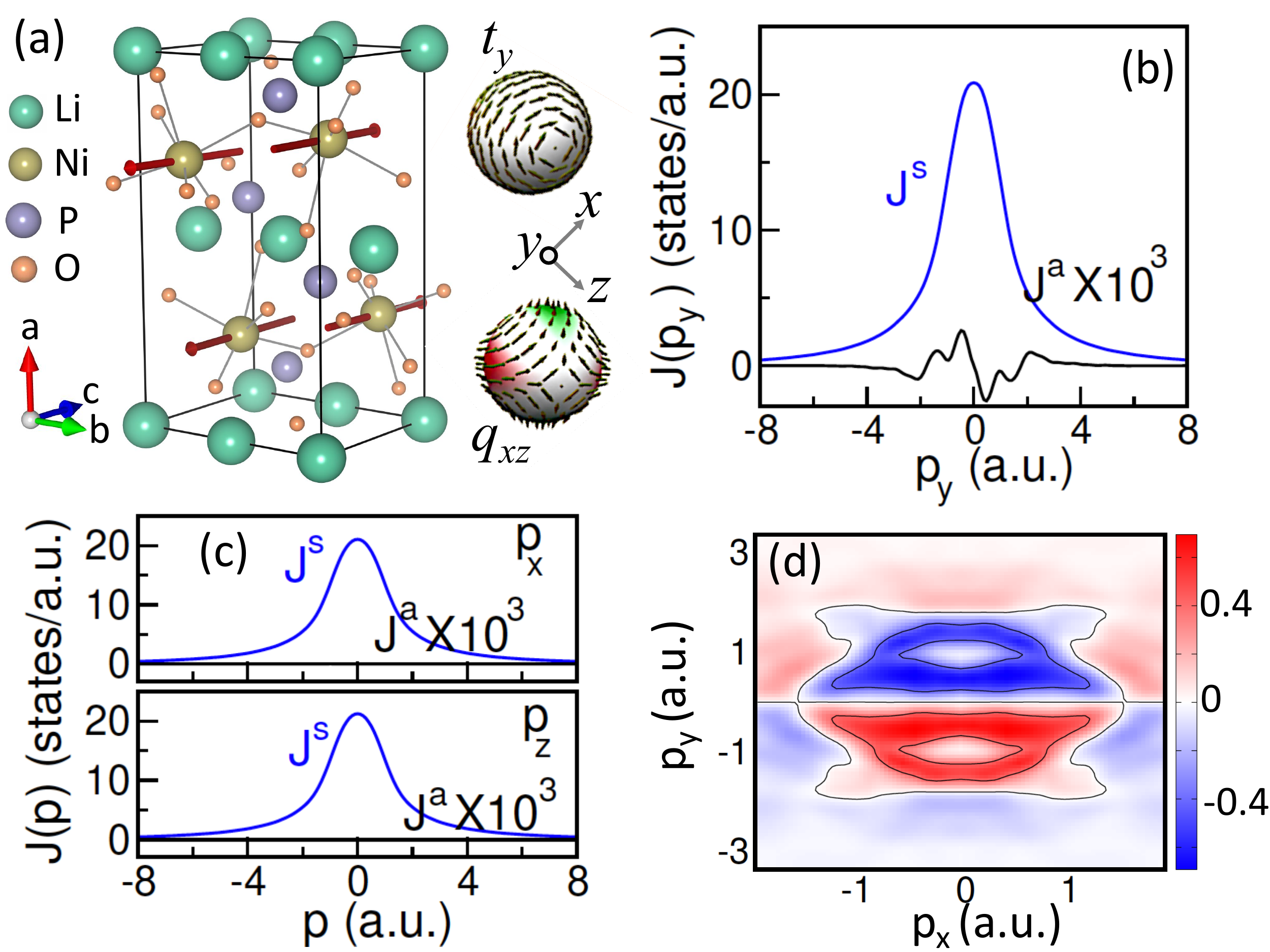}
 \caption{Anti-symmetric Compton profile in LiNiPO$_4$. (a) The crystal and magnetic ($Pnm'a$) structures  of LiNiPO$_4$. The arrows indicate the spin directions at the Ni sites. Representation of the allowed ME multipoles, toroidal moment $t_y$ and quadrupole moment $q_{xz}$, in this magnetic structure are shown at the right. (b)
 The symmetric ($J^s$) and the anti-symmetric ($J^a$) parts of the Compton profile along $y$-direction in the momentum space, computed for the magnetic structure, shown in (a). (c) The absence of 
 the anti-symmetric part for other momentum directions, $p_x$ and $p_z$. (d) The distribution of the anti-symmetric part of the line-integral of the electron momentum density along $p_z$ in the $p_x-p_y$ plane of the momentum space, indicating, further, the anti-symmetry along $p_y$. 
 The anti-symmetric parts in all figures are magnified by a factor of $10^3$.  }
 \label{fig1}
 \end{figure}

Our computed Compton profiles for the magnetic ground state along the crystallographic $x, y$, and $z$ directions [Figs. \ref{fig1} (b) and (c)] show that the anti-symmetric part  $J^a (\vec p)$ is only non-zero along the $y$ direction in  momentum space, consistent with the symmetry analysis. The symmetric and the anti-symmetric parts of the profile along $p_y$ are shown in Fig. \ref{fig1} (b). 
As seen from this figure,  both $J^s (p_y)$ and $J^a (p_y)$ satisfy the zero-sum rule $\int_{-\infty}^{\infty} p_y J^{s,a} (p_y) dp_y = 0$. While this sum rule is trivial for the symmetric part, it imposes rather a stringent constraint on each half of the antisymmetric profile, that is $\int_{0}^{\infty} p_y J^{a} (p_y) dp_y = 0$ \cite{Collins2016}. This condition indicates that for each half of $J^{a} (p_y)$, 
a positive contribution is always accompanied by a negative contribution, and vice versa, which can also be verified visually from Fig. \ref{fig1} (b). Note that the computed anti-symmetric part is rather small in magnitude, at least three orders of magnitude smaller than that of $J^s (p_y)$. The anti-symmetric part of the line integral of the electron momentum density $\int \rho^a(\vec p) dp_z$ in the $p_x-p_y$ plane is shown in Fig. \ref{fig1} (d), where it is evident that this quantity is anti-symmetric along $p_y$.  

The obtained anti-symmetric Compton profile can be related to the details of the band structure as follows: The presence or absence of $\cal I$ and $\cal T$ symmetries dictate the symmetry conditions on the band energies $E$ for any arbitrary momentum point $\vec k$ and the corresponding $-\vec k$, with
 $E (\vec k\uparrow) \overset{\cal I}{\rightarrow} E (-\vec k\uparrow)$, while $E (\vec k\uparrow) \overset{\cal T}{\rightarrow} E (-\vec k\downarrow)$. The combination of both $\cal I$ and $\cal T$ symmetries consequently leads to doubly degenerate bands at every momentum point in the BZ, $E (\vec k\uparrow) \overset{\cal IT}{\longrightarrow} E (\vec k\downarrow)$.  In the present case, individual $\cal I$ and $\cal T$ symmetries are broken but the combined ${\cal IT}$ symmetry is preserved. In this case, all bands remain doubly degenerate everywhere in the BZ due to the ${\cal IT}$ symmetry, however, there is no symmetry restriction, as the individual $\cal I$ and $\cal T$ symmetries are broken, that guarantees that bands at $\vec k$ and $-\vec k$ have the same energies. This leads to an asymmetric band structure in momentum space. 
 
\begin{figure}[t]
\centering
\includegraphics[width=\columnwidth]{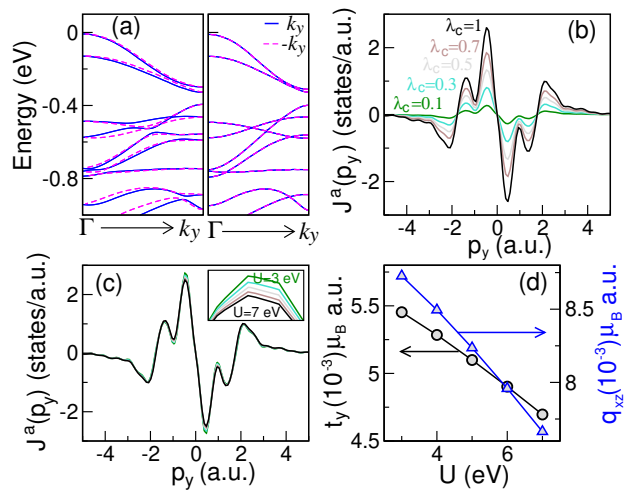}
 \caption{Role of SOC in the anti-symmetric Compton profile. (a) Band structure along $k_y$  for the magnetic configuration in Fig. \ref{fig1} (a) both in presence ({\it left}) and absence ({\it right}) of SOC. As seen from this plot, the bands are symmetric in the absence of SOC, while the presence of asymmetry is visible when SOC effects are included, indicating the crucial role of SOC in generating the asymmetry in the band dispersion. This is also reflected in the anti-symmetric Compton profile $J^a(p_y)$ in (b). (b) The variation of $J^a(p_y)$ as a function of SOC, indicating strong dependence on the scaled SOC $\lambda= \lambda_c \times \lambda_r$, where $\lambda_r$ is the strength of the SOC in the real material, and $\lambda$ is the SOC that we artificially enforce in the calculation. (c) The variation of $J^a(p_y)$ with the Hubbard $U$, indicating a weak dependence on $U$. The inset shows the small changes near the peak for a $U$ variation of 3-7 eV in intervals of 1 eV. (d) The variation of the magnitude of the computed atomic site ME multipoles $t_y$ and $q_{xz}$ with $U$. As in (c), both decrease with increasing $U$. }
 \label{fig2}
 \end{figure}

 The direction of the asymmetric band structure is determined by the momentum space basis function of the specific parity-odd, time-odd multipoles, present in the system. In the case of LiNiPO$_4$, the ME multipoles $t_y$ and $q_{xz}$ give rise to an asymmetric band structure along $k_y$, as shown in Fig. \ref{fig2} (a) ({\it left}). This asymmetric band structure, further, leads to the asymmetric Compton profile. Therefore, the asymmetry in the Compton profile, determined by the anti-symmetric part $J^{a}$, is a measure of the parity-odd, time-odd ME multipoles.
 
  It is important to point out, here, that the SOC is absolutely essential to create such an asymmetric band structure, and, hence, the asymmetric Compton profile. This is because the asymmetric magnetization density due to broken $\cal I$ and $\cal T$ symmetries, which is represented by the ME multipoles, can only couple to the space via SOC. In the absence of SOC, such coupling is absent and the band structure remains symmetric as shown in Fig. \ref{fig2} (a) ({\it right}). Since the anti-symmetric Compton profile is a manifestation of such an asymmetric band structure, it in turn also occurs only in presence of the SOC. In fact, the magnitude of $J^a (p_y)$ has a strong dependence on the strength of the SOC, as we can see by artificially changing the SOC value in LiNiPO$_4$. As depicted in Fig. \ref{fig2} (b), with decreasing the strength of the SOC constant, we find that the anti-symmetric Compton profile  decreases substantially as we decrease the strength of the SOC constant, $\lambda$.
  
  On the other hand, the dependence of $J^a (p_y)$ on the value of the Coulomb interaction parameter $U$ is quite weak, with the height of the peak in $J^a (p_y)$ decreasing only slightly with increasing $U$ [see Fig. \ref{fig2} (c)]. The weak dependence may be attributed to the slight decrease in the computed value of the ME multipoles $t_y$, and $q_{xz}$ as $U$ increases, shown in Fig. \ref{fig2} (d), indicating that the size of the anti-symmetric Compton profile depends on the size of the ME multipoles. We discuss this next.

\subsubsection{Role of ME multipoles: Effect of manipulation of the magnetic structure} \label{magStruc}

To understand the role of the ME multipoles in the Compton profile, we next explicitly compute the symmetry-allowed atomic-site ME multipoles on the Ni ions. In order to compute the atomic-site contributions, the density matrix $\rho_{lm,l'm'}$ is decomposed into tensor moments, of which the parity-odd tensor moments have contributions only from the odd $l-l'$ terms \cite{Spaldin2013}. We therefore evaluate both the $p-d$ and $s-p$ matrix element contributions. 
In particular, we manipulate the local magnetic dipolar order of the system, to cause either a change in the magnitude of the multipoles or to give rise to an entirely new set of ME multipoles. We then calculate the corresponding changes in the anti-symmetric Compton profile.

\begin{figure}[t]
\centering
\includegraphics[width=\columnwidth]{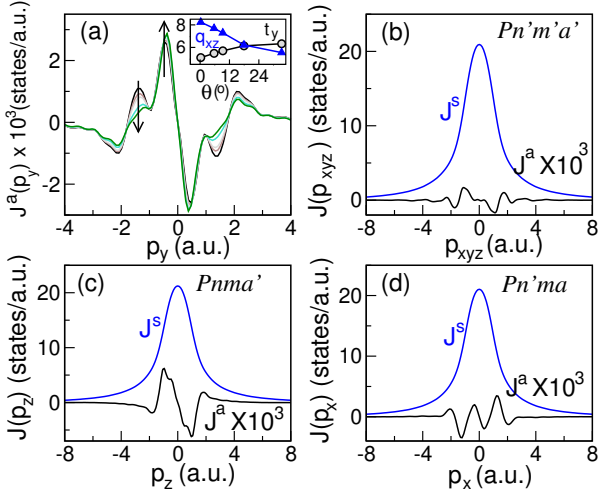}
%
 \caption{Manipulation of the magnetic structure and the corresponding anti-symmetric Compton profile. (a) The variation of the anti-symmetric Compton profile as a function of canting angle $\theta$ from $0^\circ$ to $33^\circ$. The arrow indicates the direction of increasing values of the canting angle $\theta$. The inset shows the corresponding variation in the magnitudes of the of the computed atomic-site ME toroidal moment $t_y$ and the quadrupole moment $q_{xz}$ in units of $\mu_B$ a.u. for the same values of $\theta$ as in the main plot. The symmetric ($J^s$) and the anti-symmetric ($J^a$) parts of the Compton profile along  (b) the $p_{xyz}$ direction for the magnetic structure $Pn'm'a'$, (c) $p_z$ direction for the magnetic space group $Pnma'$, and (d) $p_x$ direction for the magnetic space group $Pn'ma$. In all cases the anti-symmetric part of the Compton profile is magnified by a factor of $10^3$.}
 \label{fig3}
 \end{figure}

In the $Pnm'a$ magnetic ground state of LiNiPO$_4$, the spin moments are primarily oriented along the $z$ direction ($m_z$) with a small canting angle $\theta=\tan^{-1} (m_x/m_z)$, that gives rise to a tiny $x$ component of moment ($m_x$) as well  \cite{Jensen}. 
We investigate the effect of changing this canting angle $\theta$ on both the ME multipoles as well as the Compton profile. While changing $\theta$, the spins are always kept in the $x-z$ plane, and the $Pnm'a$ magnetic symmetry  of the structure is also kept preserved. Since this particular magnetic symmetry allows ferro-type ordering for only  $t_y$ and $q_{xz}$, no additional ferro-type ME multipoles appear as a result of change in the canting angle, but the magnitude of these multipoles may vary with the change in $\theta$. Indeed, with increase in $\theta$ the magnitude of the atomic site contributions to $t_y$ increases, while that of $q_{xz}$ decreases, as shown in the inset of Fig. \ref{fig3} (a).

Similar behavior is also seen in the local moment ({\it loc}) 
contributions to $t_y$ and $q_{xz}$, where the {\it loc} contribution to $\vec t$, $\vec t_{loc}$, can be calculated using Eq. \ref{as-lm}, and the quadrupole moment can be calculated similarly using $(q_{xz})_{loc} = 2^{-1} \sum_{\alpha} (x^\alpha m^\alpha_z+z^\alpha m^\alpha_x)$. Here $x^\alpha$, $z^\alpha$ denote the $x$ or $z$ cartesian coordinates of the Ni atom at site $\alpha$, and $m^\alpha_x$, $m^\alpha_z$ are the corresponding cartesian components of the spin moments, as listed in Table \ref{tab2}. Performing straightforward algebra, we obtain the local moment contribution to $\vec t$,
\begin{eqnarray}     \label{ME}  
\vec t_{loc} &=& 
\frac{S}{2}\Big[ \sin\theta \left(
{\begin{array}{*{20}c}
   0  \\
   4\delta c+rc\\
   qb\\
  \end{array} }  \right) + 
   \cos\theta \left(
{\begin{array}{*{20}c}
   qb  \\
   4\epsilon a+pa\\
   0\\
  \end{array} }  \right) \Big].
    \end{eqnarray} 
Here, $a, b,$ and $c$ are the orthorhombic lattice constants, $p, q,$ and $r$ are arbitrary integers, and $S$ is the magnitude of the Ni spin moment. It is easy to see that a non-trivial $\vec t_{loc}$ exists only along the $y$ direction, as it should according to symmetry, and this $y$ component increases with increasing $\theta$. In contrast, the non-trivial $(q_{xz})_{loc} = S/2 (4\delta c\sin\theta - 4\epsilon a\cos\theta)$ decreases in magnitude as $\theta$ increases. 

The corresponding changes in the anti-symmetric part of the Compton profile are depicted in Fig. \ref{fig3} (a). We see that,  with increase in $\theta$, the height of the first peak (the peak at smaller $|p_y|$) in $J^a(p_y)$ increases, whereas the second peak (the peak at larger $|p_y|$) decreases. Correlating this behavior with the simultaneous increase in $t_y$ and decrease in $q_{xz}$ calculated above, suggests that the two peaks may be associated with the two different ME multipoles $t_y$ and $q_{xz}$ respectively. 

We can, further, manipulate the magnetic structure of LiNiPO$_4$ by completely changing the spin configuration so that the symmetry changes from the actual magnetic ground state symmetry of $Pnm'a$. The introduction of  new magnetic symmetries enables different ferro-type ordered ME multipoles, offering the possibility of studying the effects of ME multipoles other than $t_y$ and $q_{xz}$ on the anti-symmetric Compton profile.

 \begin{table} [b]
\caption{The coordinates ($x^\alpha, y^\alpha, z^\alpha$) and the corresponding spin moment direction ($m^\alpha_x,m^\alpha_y,m^\alpha_z$) at each Ni site $\alpha$ corresponding to the Wyckoff positions $4c$ of the crystallographic space group $Pnma$ and the magnetic space group $Pnm'a$. $\epsilon$, $\delta$ are the internal structural parameters for LiNiPO$_4$ with orthorhombic lattice constants $a, b, c$. $S$ and $\theta$ are the magnitude of the Ni spin moment and the canting angle respectively.
}
\setlength{\tabcolsep}{3pt}
\centering
\begin{tabular}{ c c c c c c c }
\hline
Site($\alpha$)  & $x^\alpha/a$ & $y^\alpha/b$ & $z^\alpha/c$ & $m^\alpha_x$ & $m^\alpha_y$ & $m^\alpha_z$  \\
\hline\hline
Ni(1)& $-\epsilon+1/2$ & $3/4$ & $\delta+1/2$ & $S\sin\theta$ & 0 & $S\cos\theta$ \\
Ni(2) & $\epsilon$ & $1/4$ & $\delta$ & $S\sin\theta$ & 0 & $-S\cos\theta$ \\
Ni(3) & $\epsilon+1/2$ & $1/4$ & $-\delta+1/2$ & $-S\sin\theta$ & 0 & $-S\cos\theta$ \\
Ni(4) & $-\epsilon$ & $3/4$ & $-\delta$ & $-S\sin\theta$ & 0 & $S\cos\theta$ \\
  \hline
  \end{tabular}
\label{tab2}
\end{table}
\begin{figure}[t]
\centering
\includegraphics[width=\columnwidth]{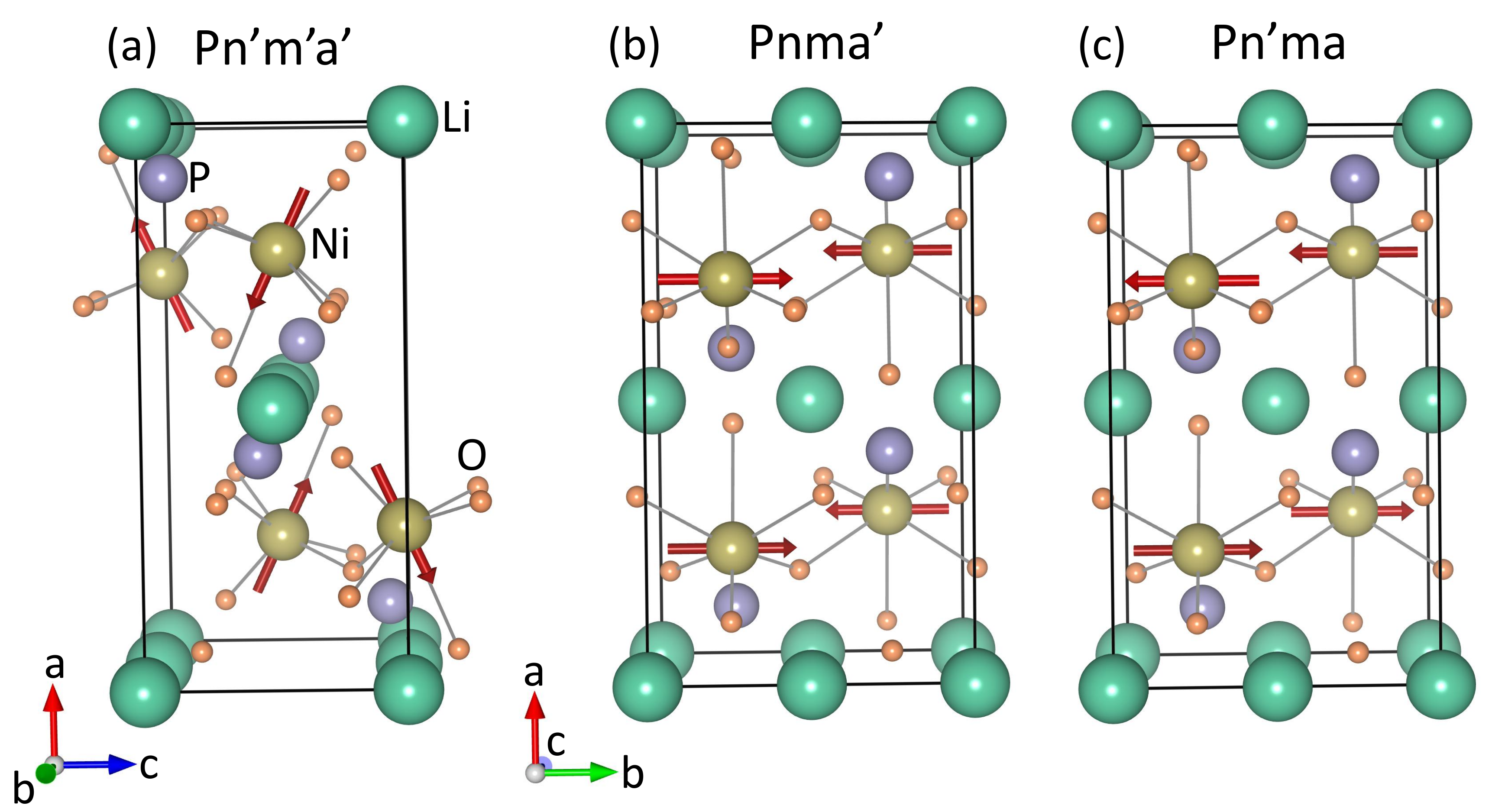}
 \caption{Schematic illustrations of the various symmetry allowed magnetic structures corresponding to the magnetic space groups (a) $Pn'm'a'$, (b) $Pnma'$, and (c) $Pn'ma$. The arrows indicate the direction of the spin moments at the Ni atoms.  }
 \label{fig6}
 \end{figure}
We made all three possible changes in the magnetic structure that are allowed by the structural $Pnma$ symmetry of LiNiPO$_4$ and also have a net ME multipole moment, while keeping the dimension of the magnetic unit cell the same as that of the structural unit cell. In the first case, we kept the Ni spin moments in the same $x-z$ plane, but changed their relative magnetic arrangement so that the $x$ components of Ni(1) and Ni(4) are parallel to each other and anti-parallel to those of Ni(2) and Ni(3), while the parallel $z$ components of Ni(1) and Ni(2) are anti-parallel to those of Ni(3) and Ni(4). This corresponds to the magnetic space group $Pn'm'a'$ [see Fig. \ref{fig6} (a)] and the IR representation $A_u^{-}$ in Table \ref{tab1}, which allows for a ME monopole $a$ and the quadrupole moments $q_{x^2-y^2}, q_{z^2}$, with the momentum space basis function $k_xk_yk_z$.
In this case, the computed Compton profile has an anti-symmetric part only along the $p_{xyz}$ direction, as shown in Fig. \ref{fig3} (b). 

We can, further, rotate the spins in the out-of-plane direction, in such a way that they have only $y$ components. Among the various possibilities, only two arrangements support a net ME multipole moment and are allowed by symmetry. These two magnetic configurations are shown in Figs. \ref{fig6} (b) and (c).
In one of these two configurations, the parallel spin moments on Ni(1), Ni(4) are anti-parallel to those of Ni(2), Ni(3), and in the other configuration Ni(1) and Ni(2) have parallel spins, which are anti-parallel to those of Ni(3) and Ni(4). The former corresponds to the magnetic space group $Pnma'$ and the IR representation $B_{1u}^{-}$, while the later corresponds to the $Pn'ma$ magnetic space group with the IR representation $B_{3u}^{-}$. Calculations for these two magnetic structures show the presence of anti-symmetric Compton profiles along $p_z$ and $p_x$ respectively [see Figs. \ref{fig3} (c) and (d)], consistent with the momentum space basis functions for the allowed multipoles $\{t_z,q_{xy} \}$ and $\{t_x,q_{yz}\}$  respectively (see Table \ref{tab1}). 

The above analysis of the manipulation of the magnetic structure provides two important conclusions. First, it confirms our argument that all the ME multipoles contribute to the anti-symmetric Compton profile and not just the toroidal moment, which is one of the central results of the present work. Secondly, it emphasizes the dependence of the anti-symmetric Compton profile on the fine details of the magnetic structure, suggesting that the manipulation of the spin configuration may also be used to decompose the profile into individual ME multipole moment contributions.

\subsection{Metallic system: ${\rm Mn_2Au}$} \label{Mn2Au}

\begin{figure}[t]
\centering
\includegraphics[width=\columnwidth]{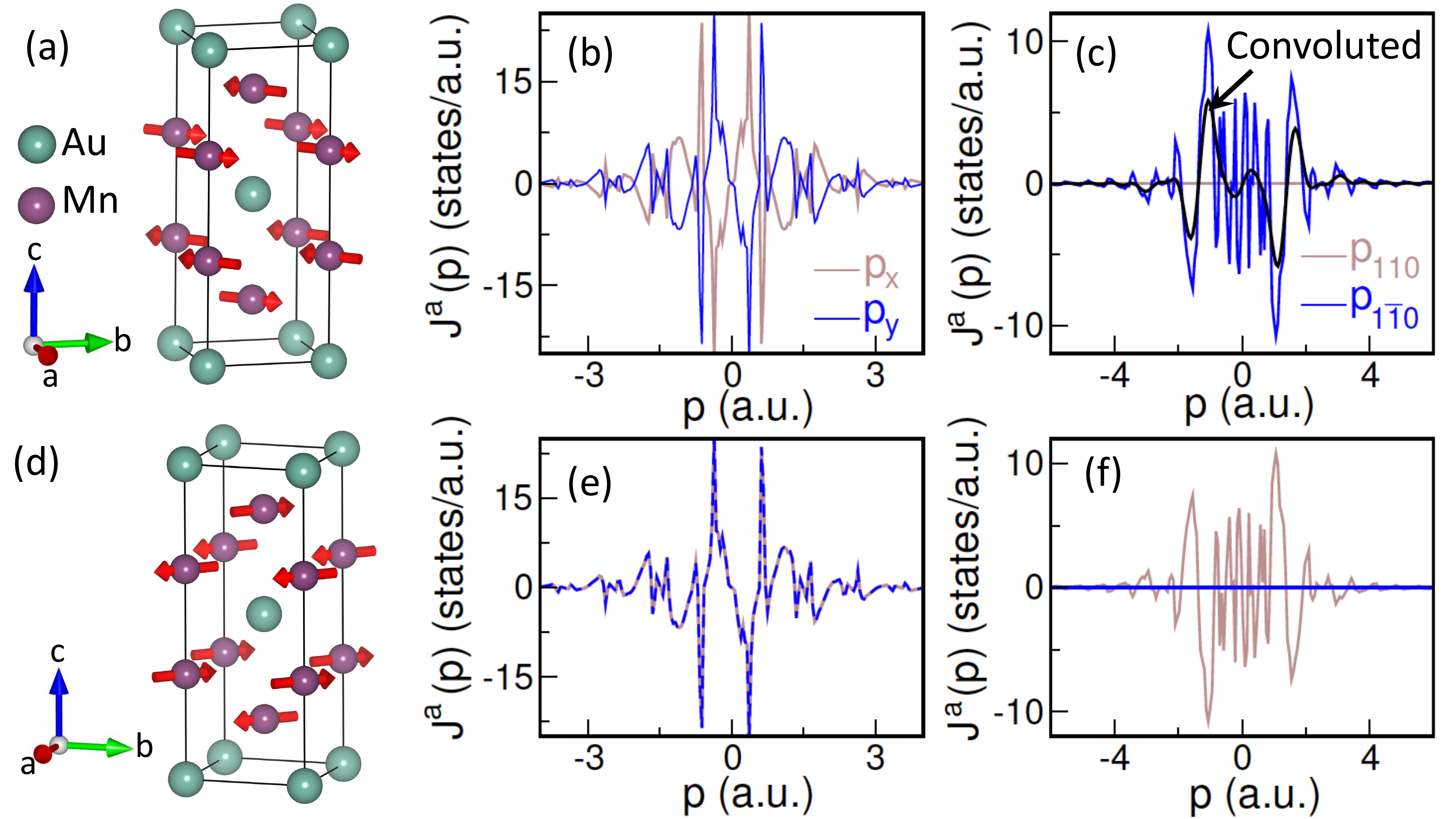}
%
 \caption{Antisymmetric Compton profile in Mn$_2$Au. (a) The crystal and magnetic structures of the ground state of Mn$_2$Au corresponding to the N\'eel vector $[110]$.  Arrows indicate the direction of the spin magnetic moments at the Mn sites. (b) 
  The anti-symmetric parts ($J^a$) of the Compton profile along $p_{x}$ and $p_y$, indicating equal and opposite contributions along these directions. This results in an asymmetry along $[1\bar{1}0]$ but an  absence of asymmetry along $[110]$. (c) The computed anti-symmetric Compton scattering profiles along the $[1\bar{1}0]$ and $[110]$ directions reflect the presence and absence of asymmetry respectively. The convolution of the 
computed anti-symmetric profile along $p_{1\bar{1}0}$ direction with a Gaussian function of 0.44 a.u. FWHM is also shown. (d) The magnetic structure for N\'eel vector $[1\bar{1}0]$. The corresponding anti-symmetric Compton profile along (e) $p_x$ and $p_y$, and (f)  $p_{110}$ and $p_{1\bar{1}0}$. In contrast to the N\'eel vector $[110]$, here the anti-symmetric profiles along $p_{x}$ and $p_y$ are the same in both magnitude and sign, leading to vanishing anti-symmetric profile along $p_{1\bar{1}0}$ but a non-zero   
$J^a$ along the $[110]$ direction.
All profiles are normalized, so that the total profile 
 is normalized to 63 electrons, which is the total number of valence electrons per formula unit of Mn$_2$Au in the calculation.   The anti-symmetric part of the Compton profile is magnified by a factor of $10^3$. }
 \label{fig4}
 \end{figure}

We now turn to our second example, Mn$_2$Au,
which crystallizes in the tetragonal structure ($I4/mmm$) with point group symmetry $D_{4h}$ \cite{Wells}. The material exhibits AFM ordering with the spins having easy $a-b$ plane anisotropy \cite{Bodnar}. 
In contrast to the previously discussed LiNiPO$_4$, Mn$_2$Au is a good conductor.  Interestingly, the electric current in Mn$_2$Au induces a switching of the N\'eel vector, which can be used to write (store) or read-out information, making it a promising material for the present day hot topic of AFM spintronics \cite{Bodnar,Chen}. The ground-state magnetic structure of Mn$_2$Au, corresponding to the magnetic space group $Fm'mm$ with $[110]$ N\'eel vector, is depicted in Fig. \ref{fig4} (a).  
Similarly to LiNiPO$_4$, the magnetic structure in Mn$_2$Au breaks the inversion symmetry, allowing for ME toroidal and quadrupole moments leading to an off-diagonal ${\cal M}_{ij}$ tensor. 
The specific magnetic configuration of Mn$_2$Au leads to interesting symmetry properties between the allowed ME multipoles, which, further, manifest in the asymmetric band structure as well as in the anti-symmetric Compton profile in Mn$_2$Au. Below we discuss these symmetry relations in detail and point out the intriguing features of the band structure and the anti-symmetric profile in Mn$_2$Au, which in addition to their relevance for Compton scattering might also be important for other experiments.

 \begin{table} [b]
\caption{The basis functions of the ME multipoles for the $D_{4h}$ point group. The symbols have the same meaning as in Table \ref{tab1}. }
\setlength{\tabcolsep}{2pt}
\centering
\begin{tabular}{ c c c   c }
\hline
IR    &  ME multipole & \multicolumn{2}{c}{Basis in}     \\
       &                        &   Real space     &  $k$-space \\ [1 ex]
\hline
 $A_{1u}^{-}$          &   $a,$   &   $xm_x+ym_y+zm_z$, &  $k_xk_yk_z (k_x^2-k_y^2)$   \\
                                &  $q_{z^2}$ & $(2zm_z-xm_x-ym_y)$ & \\
 $A_{2u}^{-} $        & $ t_z$ & $(xm_y-ym_x)$ &  $k_z$  \\
  $B_{1u}^{-}$  & $q_{x^2-y^2} $ &  $(xm_x-ym_y)$     &           $k_xk_yk_z$            \\
  $B_{2u}^{-}$  & $q_{xy} $ &  $(xm_y+ym_x)$ & $k_z(k_x^2-k_y^2)$ \\
  $E_u^-$  & $\{q_{yz},q_{xz} \}, $ & $\{ (ym_z+zm_y),  (zm_x+xm_z) \}$ & $ \{k_x,k_y \}$ \\
                   &  $\{t_x,t_y \}$ &  $\{(ym_z-zm_y),  (zm_x-xm_z)\}$ &  \\
  \hline
\end{tabular}
\label{tab3}
\end{table}
\subsubsection{Symmetry analysis}

The allowed ME multipoles and their real and momentum space basis functions in the different IR representations of the point group $D_{4h}$, relevant to Mn$_2$Au, are listed in Table \ref{tab3}. The magnetic structure of Mn$_2$Au belongs to the IR representation $E_u^-$, which allows for toroidal moments along $x$ and $y$ directions, $t_x$ and $t_y$, and also for quadrupole moments $q_{yz}$ and $q_{xz}$. No additional local ME multipoles are allowed at the Mn site, forbidding any additional ME multipoles with anti-ferro type arrangement. It is interesting to point out here that the allowed multipole tensor for the magnetic ground state, corresponding to the magnetic space group $Fm'mm$, has the constraints ${\cal M}_{xz} = {\cal M}_{yz}$, and ${\cal M}_{zx} = {\cal M}_{zy}$.
 This, in turn, imposes an additional constraint on the allowed ME multipoles, that $t_x+t_y =q_{yz}-q_{xz}$ (see Eq. \ref{Mtensor}).

As expected, the momentum-space basis functions of the time-odd, parity-odd ME multipoles in Table \ref{tab3} are always spin independent and odd order in $k$.
According to the momentum-space basis function for the allowed ME multipoles in Mn$_2$Au, anti-symmetric Compton profiles should appear along $k_x$ and $k_y$ directions (see Table \ref{tab3}), which we, further, verify by explicitly computing the anti-symmetric Compton profile  as discussed below.

\subsubsection{DFT results for the Compton profile and ME multipoles}

We compute the anti-symmetric Compton profile for Mn$_2$Au using the ELK  code \cite{code,elk}, following the same method  as discussed in section \ref{DFT-LNPO}. All calculations are performed with the experimentally reported structure \cite{Wells} within the LDA+SOC+$U$ formalism with the correlation parameters $U= 1$ eV and $J= 0.5$ eV at the Mn site. The specific choice of the correlation parameters is guided by recent photoemission measurements \cite{Elmers}, the results of which can be reproduced satisfactorily with these parameters.
 It is important to point out here that, in contrast to the insulating LiNiPO$_4$, the anti-symmetric Compton profile in Mn$_2$Au has a strong dependence on the choice of $k$-point mesh, and the convergence of the anti-symmetric profile is achieved at a much higher $k$-point grid of $31\times31\times12$. 

In the computed magnetic ground state, the magnetic moment at each Mn site projected onto  ionic radius 2.4 a.u. is about 4.1 $\mu_B$. In this magnetic structure, our
calculations find that
 only the $t_x$, $t_y$, $q_{xz}$, and $q_{yz}$ ME multipoles have non-zero values, and that they have ferro-type arrangements between different Mn atoms, as expected from the symmetry analysis. In addition, the computed components of the toroidal moment as well as the quadrupole moments are not independent but are related to each other by $t_x=-t_y$, and $q_{xz}=q_{yz}$. These relations satisfy the symmetry imposed condition on the ME multipoles $t_x+t_y =q_{yz}-q_{xz}$, as mentioned earlier. 
 
 Interestingly, for the $[1\bar{1}0]$ N\'eel vector orientation, which is symmetry equivalent to the $[110]$ N\'eel vector but represents a different magnetoelectric domain \cite{Elmers,Sapozhnik}, the relative signs of the toroidal and the quadrupole moment components switch, such that $t_x=t_y$, and $q_{xz}= -q_{yz}$. 
  These relative orientations of the computed multipoles can be understood directly from their formal definitions, given in section \ref{sec1}. For example, according to the definition of the toroidal moment, $t_x=\frac{1}{2} (ym_z-zm_y)$ and $t_y=\frac{1}{2} (xm_z-zm_x)$,  which gives $t_x=-t_y$ for $m_x=m_y$ and $m_z=0$ ($[110]$ N\'eel vector), but $t_x= t_y$ for $m_x=-m_y$ and $m_z=0$ ($[1\bar{1}0]$ N\'eel vector). Similarly, it is easy to show that $q_{xz}$ and $q_{yz}$ are the same in both sign and magnitude for $[110]$ orientation of the N\'eel vector, while they have opposite signs when the N\'eel vector is along $[1\bar{1}0]$.
The computed magnitudes of the independent components of the allowed ME multipoles at the Mn site are $t_y=-5.3 \times 10^{-3} \mu_B$ a.u. and $q_{xz} = -6.2  \times 10^{-3} \mu_B$ a.u., where we have added the $p-d$ and $p-s$ matrix element contributions to obtain the total atomic site contribution for each Mn atom. These ME multipoles lead to interesting consequences in the corresponding band structure and the anti-symmetric Compton profile as we now proceed to discuss.

The calculation of the Compton profile along the three Cartesian directions $x$, $y$, and $z$ shows 
an anti-symmetric component only along $p_x$ and $p_y$, while $J^a$ is absent along $p_z$, in agreement with the symmetry analysis, discussed earlier. As seen from Figs. \ref{fig4} (b) and (c),  unlike LiNiPO$_4$, the anti-symmetric profile in Mn$_2$Au has sharp peaks at lower momentum, which also change their signs rapidly. 
Also, the anti-symmetric profile has a larger magnitude compared to that in LiNiPO$_4$, which may be attributed to the presence of the heavy Au atom leading to stronger SOC effects in Mn$_2$Au.

It is interesting to point out here that the anti-symmetric parts along $p_x$ and $p_y$ directions are exactly equal and opposite to each other, i.e., $J^a(p_x) =-J^a(p_y)$ [see Fig. \ref{fig4} (b)]. This leads to a vanishing anti-symmetric Compton profile along the $[110]$ direction in momentum space, while a non-zero anti-symmetric profile appears in the perpendicular direction $[1\bar{1}0]$ as shown in Fig. \ref{fig4} (c).
The convolution of the computed profile with an experimental resolution (Gaussian of FWHM) of   0.44 a.u. shows that while the sharp peaks as well as their oscillations in low momentum regime may be suppressed, some oscillating features can still be captured within this resolution limit [see Fig. \ref{fig4} (c)].  
 
 Furthermore, switching of the N\'eel vector to the $[1\bar{1}0]$ direction results in an exactly opposite situation, as depicted in Figs. \ref{fig4} (e) and (f). In this case, the anti-symmetric profiles along $p_x$ and $p_y$ are equal in both magnitude and sign, leading to the presence of an anti-symmetric Compton profile along $[110]$ direction in momentum space, while it is absent along $[1\bar{1}0]$. 
This may be of importance to the experimental measurements, where an electric current-induced switching of the N\'eel vector, in turn, results in a switching of the anti-symmetric Compton profile from the $[1\bar{1}0]$ to the $[110]$ direction in the momentum space. 

\begin{figure}[t]
\centering
\includegraphics[width=\columnwidth]{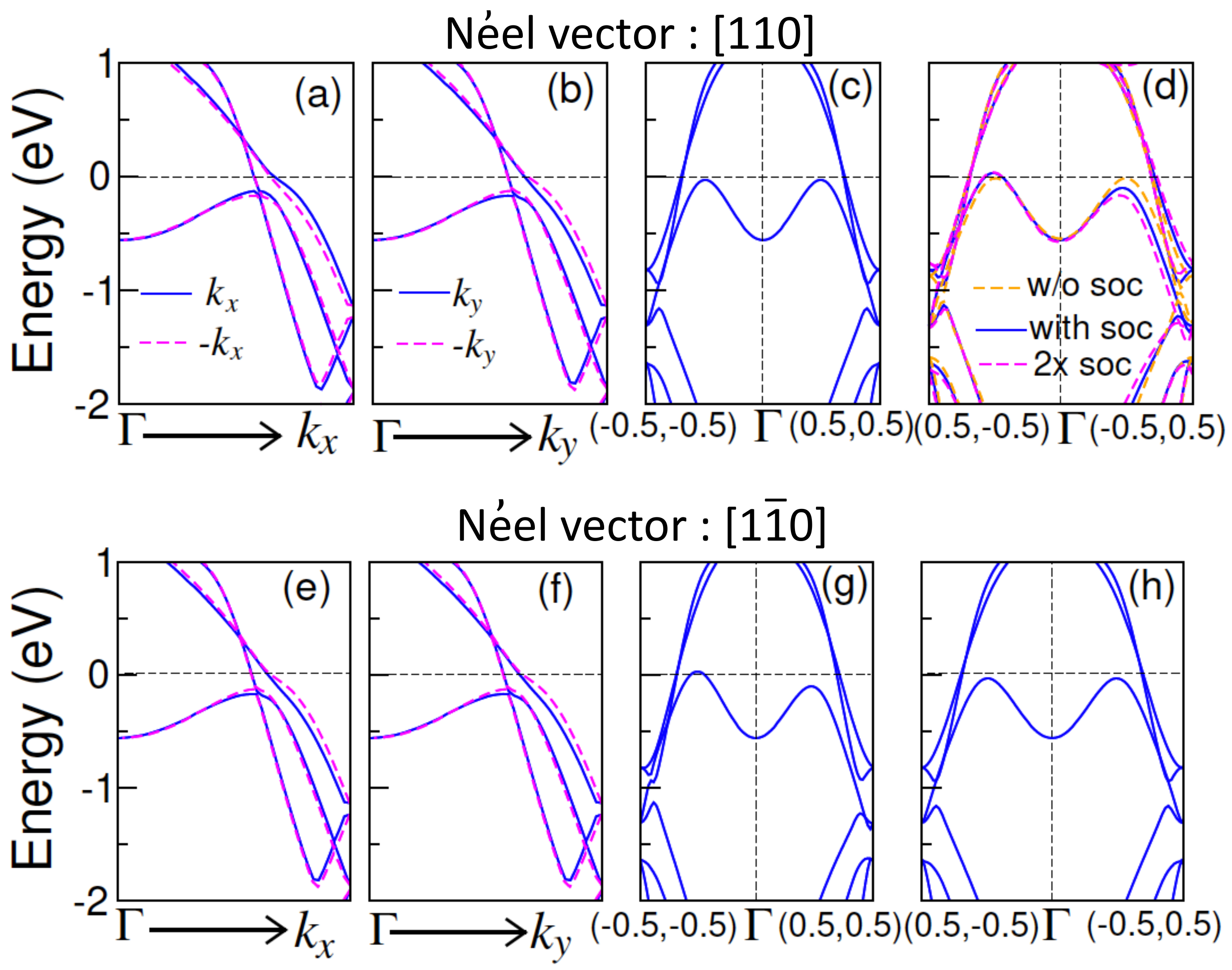}
%
 \caption{The asymmetry in the band structure of Mn$_2$Au, induced by the ME multipoles, for the N\'eel vectors $[110]$ (top panel) and $[1\bar{1}0]$ (bottom panel). The band structure along (a) $k_x$, (b) $k_y$, (c) $[110]$, and (d) $[1\bar{1}0]$ directions in the momentum space for the N\'eel vector $[110]$ with the corresponding magnetic structure shown in Fig. \ref{fig4} (a). The comparison of (a) and (b) shows that the asymmetry of the bands are opposite along $k_x$ and $k_y$ for the N\'eel vector $[110]$. This results in an absence of asymmetry along $[110]$ direction in (c), while the asymmetry along $[1\bar{1}0]$ direction is clearly visible in (d).   
 Comparison with bands in absence of SOC shows that the asymmetry is present only in presence of SOC. Doubling the strength of the SOC further enhances this asymmetry, indicating the importance of SOC. (e)-(h) The same as in the top panel but for the  N\'eel vector $[1\bar{1}0]$. The corresponding magnetic structure is in Fig. \ref{fig4} (d). In this case bands along (e) $k_x$ and (f) $k_y$ have the same asymmetry, leading to asymmetric bands along (g) $[110]$ direction, and absence of asymmetry along (h)  $[1\bar{1}0]$ directions in the momentum space.
 }
 \label{fig5}
 \end{figure}

These distinct features of the anti-symmetric Compton profile in Mn$_2$Au may be understood from careful analysis of the corresponding band structures. The computed band structure for the magnetic ground state of Mn$_2$Au with $[110]$ N\'eel vector shows that the bands are asymmetric  along $k_x$ and $k_y$ directions [see Figs. \ref{fig5} (a) and (b)]. The asymmetry of the bands in the presence of SOC follows from similar symmetry arguments to those discussed above for the case of LiNiPO$_4$. Interestingly, the comparison of the bands along $k_x$ and $k_y$ shows that the band asymmetry is opposite for these two directions, explaining the opposite signs of the computed anti-symmetric profiles along $p_x$ and $p_y$. The asymmetry of the bands is even more pronounced along the $[1\bar{1}0]$ direction of the momentum space, while along the $[110]$ direction the bands are absolutely symmetric as depicted in Figs. \ref{fig5} (c) and (d). This explains, further, the presence and the absence of the anti-symmetric Compton profile along the $[1\bar{1}0]$ and $[110]$ directions respectively.

In contrast, switching of the magnetization direction to $[1\bar{1}0]$ results in similar asymmetries along $k_x$ and $k_y$ directions as understood by comparing the bands shown in Figs.  \ref{fig5} (e) and (f). This results in a band asymmetry along the $[110]$ direction, while the bands remain symmetric in the perpendicular $[1\bar{1}0]$ direction [see Figs. \ref{fig5} (g) and (h)], consistent with the switching of the anti-symmetric Compton profile by changing the orientation of the N\'eel vector in Mn$_2$Au. Interestingly, changes in the energy of the conduction electrons corresponding to the different N\'eel vectors are also recently reported in the photo-emission measurements on Mn$_2$Au thin films \cite{Elmers}. (Note that the unit cell of the reported thin film in Ref.~\onlinecite{Elmers} is rotated with respect to the bulk unit cell considered in our calculation, 
and our $[1\bar{1}0]$ and $[110]$ directions correspond to their $k_x$, $k_y$ directions respectively.) The confirmation of left-right asymmetry in the band structure is, however, not possible from these measurements as the photo-emission intensity from the two domains was averaged out \cite{Elmers}. 

Similarly to the case of LiNiPO$_4$, these asymmetries in the band structure can be traced back to the allowed ME multipoles in momentum space (see Table \ref{tab3}). Physically speaking, the spin asymmetry, designated by the allowed ME multipoles, couples to the momentum space via SOC, resulting in a band asymmetry. This is further evidenced by comparing the band structures for two different values of SOC, as shown in Fig. \ref{fig5} (d), where we see that doubling the strength of the SOC results in an enhanced asymmetry.

Finally, we want to emphasize the one-to-one correspondence between the band asymmetries along $[1\bar{1}0]$ or $[110]$ directions in the momentum space for the two N\'eel vectors $[110]$ and $[1\bar{1}0]$ in Mn$_2$Au and the respective directions of the allowed toroidal moment vectors  $\vec t = t_y (\hat i -\hat j)$ or $\vec t = t_y (\hat i + \hat j)$. 
We note that the components of the quadrupole moment tensors $q_{xz}, q_{yz}$ have also the same symmetry as the allowed toroidal moment, belonging to the same $E_u^-$ irreducible representation.
It is interesting to point out that the same ME multipoles can provide a convenient description of the 
current induced switching of the N\'eel vectors in Mn$_2$Au \cite{Florian}.  
   Our theoretical analysis suggests a presence of band asymmetry in Mn$_2$Au and its switching with the change in N\'eel vector, that are  signatures of the allowed ME multipole moments in the system.

\section{Summary and outlook}  \label{sec3}

In summary, we have presented a comprehensive theoretical analysis of the anti-symmetric Compton profile in materials with non-centrosymmetric magnetic ordering. We have shown that the anti-symmetric Compton profile is a   signature of parity-odd, time-odd ME multipoles. 
The ME multipoles, which break both space-inversion and time-reversal symmetries, generate an asymmetry in the magnetization density, the details of which are governed by the specific components of the ME multipoles that occur. The asymmetry, in turn,  can be characterized by  the basis functions of the corresponding ME multipoles, which, since the ME multipoles are odd parity, are different in real and momentum space. In particular, while the real space basis functions for ME multipoles are always spin dependent, the momentum space basis functions are not. As a result, they can be probed in $k$-space without any spin-resolved measurements. In addition, since the momentum-space bases for ME multipoles are always odd order in $k$, they generate asymmetry in the electronic band dispersion in momentum space, with the direction of asymmetry  determined by the specific momentum-space basis. Since the Compton scattering is a measure of the electron-momentum density, an asymmetry in the Compton scattering profile is generated in the presence of ME multipoles. The coupling of the magnetization asymmetry to the space requires the presence of SOC, indicating the importance of SOC effects in generating the anti-symmetric Compton profile.

In addition to emphasizing the role of the real space-momentum space duality in understanding the anti-symmetric profile, our work points out the two essential ingredients for an anti-symmetric Compton profile: time-odd, parity-odd ME multipoles and SOC. This can in turn be used to guide the search for suitable candidate materials with larger anti-symmetric Compton profiles. We note that earlier attempts were not inconsistent with the theoretical predictions, but were not conclusive as the observed anti-symmetric signal fell within the statistical error \cite{Collins2016,LNPO}.
The ME multipoles requirement means that the anti-symmetric  profile has a strong dependence on the magnetic structure of the system. This leads to the possibility of tuning the profile by manipulating the magnetic structure, as discussed in section \ref{magStruc}. The SOC requirement suggests a route to identifying materials with a stronger effect by choosing materials with heavy elements, and therefore strong SOC.  Furthermore, the finding of the current work that the anti-symmetric Compton profile is present for all ME materials, and not just those with ME toroidal moment, also broadens the scope for candidate materials.

Finally, our theoretical studies on the representative insulating and metallic candidate materials indicate some distinct features in the corresponding anti-symmetric Compton profiles. While the anti-symmetric profile has broad peaks in  insulating LiNiPO$_4$, the peaks are rather sharp and oscillating, specially in the low momentum regime, for the metallic system Mn$_2$Au. We also predict some detailed features that should be experimentally observable, for example opposite changes in the two peak heights in LiNiPO$_4$ as the spin canting angle is varied, and may give further insight into the detailed role of the ME multipoles. Of particular interest is the anti-symmetric Compton profile in Mn$_2$Au, the direction of which can be switched between $p_{1\bar{1}0}$ and $p_{110}$ by changing the orientation of the Mn spin moments from $[110]$ to $[1\bar{1}0]$. We hope that our predictions will stimulate more experiments in this direction.

\section*{Acknowledgements}
The authors thank Stephen Collins, Jon Duffy, Michael Fechner, and Urs Staub for stimulating discussions. 
This work was supported by the European Research Council (ERC) under the European Union’s Horizon 2020 research and innovation programme project HERO grant agreement No. 810451. Computational resources were provided by ETH Z\"urich (Euler cluster).


\bibliographystyle{apsrev4-1}
\bibliography{LNPO}

\begin{thebibliography}{40}%
\makeatletter
\providecommand \@ifxundefined [1]{%
 \@ifx{#1\undefined}
}%
\providecommand \@ifnum [1]{%
 \ifnum #1\expandafter \@firstoftwo
 \else \expandafter \@secondoftwo
 \fi
}%
\providecommand \@ifx [1]{%
 \ifx #1\expandafter \@firstoftwo
 \else \expandafter \@secondoftwo
 \fi
}%
\providecommand \natexlab [1]{#1}%
\providecommand \enquote  [1]{``#1''}%
\providecommand \bibnamefont  [1]{#1}%
\providecommand \bibfnamefont [1]{#1}%
\providecommand \citenamefont [1]{#1}%
\providecommand \href@noop [0]{\@secondoftwo}%
\providecommand \href [0]{\begingroup \@sanitize@url \@href}%
\providecommand \@href[1]{\@@startlink{#1}\@@href}%
\providecommand \@@href[1]{\endgroup#1\@@endlink}%
\providecommand \@sanitize@url [0]{\catcode `\\12\catcode `\$12\catcode
  `\&12\catcode `\#12\catcode `\^12\catcode `\_12\catcode `\%12\relax}%
\providecommand \@@startlink[1]{}%
\providecommand \@@endlink[0]{}%
\providecommand \url  [0]{\begingroup\@sanitize@url \@url }%
\providecommand \@url [1]{\endgroup\@href {#1}{\urlprefix }}%
\providecommand \urlprefix  [0]{URL }%
\providecommand \Eprint [0]{\href }%
\providecommand \doibase [0]{http://dx.doi.org/}%
\providecommand \selectlanguage [0]{\@gobble}%
\providecommand \bibinfo  [0]{\@secondoftwo}%
\providecommand \bibfield  [0]{\@secondoftwo}%
\providecommand \translation [1]{[#1]}%
\providecommand \BibitemOpen [0]{}%
\providecommand \bibitemStop [0]{}%
\providecommand \bibitemNoStop [0]{.\EOS\space}%
\providecommand \EOS [0]{\spacefactor3000\relax}%
\providecommand \BibitemShut  [1]{\csname bibitem#1\endcsname}%
\let\auto@bib@innerbib\@empty
\bibitem [{\citenamefont {Landau}\ and\ \citenamefont
  {Lifshitz}(1980)}]{LandauLifshitz}%
  \BibitemOpen
  \bibfield  {author} {\bibinfo {author} {\bibfnamefont {L.~D.}\ \bibnamefont
  {Landau}}\ and\ \bibinfo {author} {\bibfnamefont {E.~M.}\ \bibnamefont
  {Lifshitz}},\ }\href@noop {} {\emph {\bibinfo {title} {The Classical Theory
  of Fields}}}\ (\bibinfo  {publisher} {Butterworth-Heinemann, Oxford (4th
  ed)},\ \bibinfo {year} {1980})\BibitemShut {NoStop}%
\bibitem [{\citenamefont {Jackson}(1999)}]{Jackson}%
  \BibitemOpen
  \bibfield  {author} {\bibinfo {author} {\bibfnamefont {J.~D.}\ \bibnamefont
  {Jackson}},\ }\href@noop {} {\emph {\bibinfo {title} {Classical
  Electrodynamics}}}\ (\bibinfo  {publisher} {Wiley, New York (3rd ed.)},\
  \bibinfo {year} {1999})\BibitemShut {NoStop}%
\bibitem [{\citenamefont {Blin-Stoyle}(1956)}]{Stoyle}%
  \BibitemOpen
  \bibfield  {author} {\bibinfo {author} {\bibfnamefont {R.~J.}\ \bibnamefont
  {Blin-Stoyle}},\ }\href {\doibase 10.1103/RevModPhys.28.75} {\bibfield
  {journal} {\bibinfo  {journal} {Rev. Mod. Phys.}\ }\textbf {\bibinfo {volume}
  {28}},\ \bibinfo {pages} {75} (\bibinfo {year} {1956})}\BibitemShut {NoStop}%
\bibitem [{\citenamefont {Raab}\ and\ \citenamefont {Lange}(2004)}]{Raab}%
  \BibitemOpen
  \bibfield  {author} {\bibinfo {author} {\bibfnamefont {R.~E.}\ \bibnamefont
  {Raab}}\ and\ \bibinfo {author} {\bibfnamefont {O.~L.~D.}\ \bibnamefont
  {Lange}},\ }\href@noop {} {\emph {\bibinfo {title} {Multipole Theory in
  Electromagnetism: Classical, Quantum, and Symmetry Aspects, with
  Applications}}}\ (\bibinfo  {publisher} {Oxford University Press, Oxford,
  U.K.},\ \bibinfo {year} {2004})\BibitemShut {NoStop}%
\bibitem [{\citenamefont {Kaelberer}\ \emph {et~al.}(2010)\citenamefont
  {Kaelberer}, \citenamefont {Fedotov}, \citenamefont {Papasimakis},
  \citenamefont {Tsai},\ and\ \citenamefont {Zheludev}}]{Kaelberer}%
  \BibitemOpen
  \bibfield  {author} {\bibinfo {author} {\bibfnamefont {T.}~\bibnamefont
  {Kaelberer}}, \bibinfo {author} {\bibfnamefont {V.~A.}\ \bibnamefont
  {Fedotov}}, \bibinfo {author} {\bibfnamefont {N.}~\bibnamefont
  {Papasimakis}}, \bibinfo {author} {\bibfnamefont {D.~P.}\ \bibnamefont
  {Tsai}}, \ and\ \bibinfo {author} {\bibfnamefont {N.~I.}\ \bibnamefont
  {Zheludev}},\ }\href {\doibase 10.1126/science.1197172} {\bibfield  {journal}
  {\bibinfo  {journal} {Science}\ }\textbf {\bibinfo {volume} {330}},\ \bibinfo
  {pages} {1510} (\bibinfo {year} {2010})}\BibitemShut {NoStop}%
\bibitem [{\citenamefont {Spaldin}\ \emph {et~al.}(2008)\citenamefont
  {Spaldin}, \citenamefont {Fiebig},\ and\ \citenamefont
  {Mostovoy}}]{Spaldin2008}%
  \BibitemOpen
  \bibfield  {author} {\bibinfo {author} {\bibfnamefont {N.~A.}\ \bibnamefont
  {Spaldin}}, \bibinfo {author} {\bibfnamefont {M.}~\bibnamefont {Fiebig}}, \
  and\ \bibinfo {author} {\bibfnamefont {M.}~\bibnamefont {Mostovoy}},\ }\href
  {\doibase 10.1088/0953-8984/20/43/434203} {\bibfield  {journal} {\bibinfo
  {journal} {Journal of Physics: Condensed Matter}\ }\textbf {\bibinfo {volume}
  {20}},\ \bibinfo {pages} {434203} (\bibinfo {year} {2008})}\BibitemShut
  {NoStop}%
\bibitem [{\citenamefont {Hayami}\ \emph {et~al.}(2018)\citenamefont {Hayami},
  \citenamefont {Yatsushiro}, \citenamefont {Yanagi},\ and\ \citenamefont
  {Kusunose}}]{Hayami}%
  \BibitemOpen
  \bibfield  {author} {\bibinfo {author} {\bibfnamefont {S.}~\bibnamefont
  {Hayami}}, \bibinfo {author} {\bibfnamefont {M.}~\bibnamefont {Yatsushiro}},
  \bibinfo {author} {\bibfnamefont {Y.}~\bibnamefont {Yanagi}}, \ and\ \bibinfo
  {author} {\bibfnamefont {H.}~\bibnamefont {Kusunose}},\ }\href {\doibase
  10.1103/PhysRevB.98.165110} {\bibfield  {journal} {\bibinfo  {journal} {Phys.
  Rev. B}\ }\textbf {\bibinfo {volume} {98}},\ \bibinfo {pages} {165110}
  (\bibinfo {year} {2018})}\BibitemShut {NoStop}%
\bibitem [{\citenamefont {Spaldin}\ \emph {et~al.}(2013)\citenamefont
  {Spaldin}, \citenamefont {Fechner}, \citenamefont {Bousquet}, \citenamefont
  {Balatsky},\ and\ \citenamefont {Nordstr\"om}}]{Spaldin2013}%
  \BibitemOpen
  \bibfield  {author} {\bibinfo {author} {\bibfnamefont {N.~A.}\ \bibnamefont
  {Spaldin}}, \bibinfo {author} {\bibfnamefont {M.}~\bibnamefont {Fechner}},
  \bibinfo {author} {\bibfnamefont {E.}~\bibnamefont {Bousquet}}, \bibinfo
  {author} {\bibfnamefont {A.}~\bibnamefont {Balatsky}}, \ and\ \bibinfo
  {author} {\bibfnamefont {L.}~\bibnamefont {Nordstr\"om}},\ }\href {\doibase
  10.1103/PhysRevB.88.094429} {\bibfield  {journal} {\bibinfo  {journal} {Phys.
  Rev. B}\ }\textbf {\bibinfo {volume} {88}},\ \bibinfo {pages} {094429}
  (\bibinfo {year} {2013})}\BibitemShut {NoStop}%
\bibitem [{\citenamefont {Watanabe}\ and\ \citenamefont
  {Yanase}(2020)}]{Watanabe2020}%
  \BibitemOpen
  \bibfield  {author} {\bibinfo {author} {\bibfnamefont {H.}~\bibnamefont
  {Watanabe}}\ and\ \bibinfo {author} {\bibfnamefont {Y.}~\bibnamefont
  {Yanase}},\ }\href {\doibase 10.1103/PhysRevResearch.2.043081} {\bibfield
  {journal} {\bibinfo  {journal} {Phys. Rev. Research}\ }\textbf {\bibinfo
  {volume} {2}},\ \bibinfo {pages} {043081} (\bibinfo {year}
  {2020})}\BibitemShut {NoStop}%
\bibitem [{\citenamefont {Dubovik}\ and\ \citenamefont
  {Tugushev}(1990)}]{Dubovik}%
  \BibitemOpen
  \bibfield  {author} {\bibinfo {author} {\bibfnamefont {V.}~\bibnamefont
  {Dubovik}}\ and\ \bibinfo {author} {\bibfnamefont {V.}~\bibnamefont
  {Tugushev}},\ }\href {\doibase https://doi.org/10.1016/0370-1573(90)90042-Z}
  {\bibfield  {journal} {\bibinfo  {journal} {Physics Reports}\ }\textbf
  {\bibinfo {volume} {187}},\ \bibinfo {pages} {145} (\bibinfo {year}
  {1990})}\BibitemShut {NoStop}%
\bibitem [{\citenamefont {Onimaru}\ \emph {et~al.}(2011)\citenamefont
  {Onimaru}, \citenamefont {Matsumoto}, \citenamefont {Inoue}, \citenamefont
  {Umeo}, \citenamefont {Sakakibara}, \citenamefont {Karaki}, \citenamefont
  {Kubota},\ and\ \citenamefont {Takabatake}}]{Onimaru}%
  \BibitemOpen
  \bibfield  {author} {\bibinfo {author} {\bibfnamefont {T.}~\bibnamefont
  {Onimaru}}, \bibinfo {author} {\bibfnamefont {K.~T.}\ \bibnamefont
  {Matsumoto}}, \bibinfo {author} {\bibfnamefont {Y.~F.}\ \bibnamefont
  {Inoue}}, \bibinfo {author} {\bibfnamefont {K.}~\bibnamefont {Umeo}},
  \bibinfo {author} {\bibfnamefont {T.}~\bibnamefont {Sakakibara}}, \bibinfo
  {author} {\bibfnamefont {Y.}~\bibnamefont {Karaki}}, \bibinfo {author}
  {\bibfnamefont {M.}~\bibnamefont {Kubota}}, \ and\ \bibinfo {author}
  {\bibfnamefont {T.}~\bibnamefont {Takabatake}},\ }\href {\doibase
  10.1103/PhysRevLett.106.177001} {\bibfield  {journal} {\bibinfo  {journal}
  {Phys. Rev. Lett.}\ }\textbf {\bibinfo {volume} {106}},\ \bibinfo {pages}
  {177001} (\bibinfo {year} {2011})}\BibitemShut {NoStop}%
\bibitem [{\citenamefont {Cricchio}\ \emph {et~al.}(2009)\citenamefont
  {Cricchio}, \citenamefont {Bultmark}, \citenamefont {Gr\aa{}n\"as},\ and\
  \citenamefont {Nordstr\"om}}]{Cricchio}%
  \BibitemOpen
  \bibfield  {author} {\bibinfo {author} {\bibfnamefont {F.}~\bibnamefont
  {Cricchio}}, \bibinfo {author} {\bibfnamefont {F.}~\bibnamefont {Bultmark}},
  \bibinfo {author} {\bibfnamefont {O.}~\bibnamefont {Gr\aa{}n\"as}}, \ and\
  \bibinfo {author} {\bibfnamefont {L.}~\bibnamefont {Nordstr\"om}},\ }\href
  {\doibase 10.1103/PhysRevLett.103.107202} {\bibfield  {journal} {\bibinfo
  {journal} {Phys. Rev. Lett.}\ }\textbf {\bibinfo {volume} {103}},\ \bibinfo
  {pages} {107202} (\bibinfo {year} {2009})}\BibitemShut {NoStop}%
\bibitem [{\citenamefont {Santini}\ \emph {et~al.}(2009)\citenamefont
  {Santini}, \citenamefont {Carretta}, \citenamefont {Amoretti}, \citenamefont
  {Caciuffo}, \citenamefont {Magnani},\ and\ \citenamefont {Lander}}]{Santini}%
  \BibitemOpen
  \bibfield  {author} {\bibinfo {author} {\bibfnamefont {P.}~\bibnamefont
  {Santini}}, \bibinfo {author} {\bibfnamefont {S.}~\bibnamefont {Carretta}},
  \bibinfo {author} {\bibfnamefont {G.}~\bibnamefont {Amoretti}}, \bibinfo
  {author} {\bibfnamefont {R.}~\bibnamefont {Caciuffo}}, \bibinfo {author}
  {\bibfnamefont {N.}~\bibnamefont {Magnani}}, \ and\ \bibinfo {author}
  {\bibfnamefont {G.~H.}\ \bibnamefont {Lander}},\ }\href {\doibase
  10.1103/RevModPhys.81.807} {\bibfield  {journal} {\bibinfo  {journal} {Rev.
  Mod. Phys.}\ }\textbf {\bibinfo {volume} {81}},\ \bibinfo {pages} {807}
  (\bibinfo {year} {2009})}\BibitemShut {NoStop}%
\bibitem [{\citenamefont {Sumita}\ \emph {et~al.}(2017)\citenamefont {Sumita},
  \citenamefont {Nomoto},\ and\ \citenamefont {Yanase}}]{Sumita}%
  \BibitemOpen
  \bibfield  {author} {\bibinfo {author} {\bibfnamefont {S.}~\bibnamefont
  {Sumita}}, \bibinfo {author} {\bibfnamefont {T.}~\bibnamefont {Nomoto}}, \
  and\ \bibinfo {author} {\bibfnamefont {Y.}~\bibnamefont {Yanase}},\ }\href
  {\doibase 10.1103/PhysRevLett.119.027001} {\bibfield  {journal} {\bibinfo
  {journal} {Phys. Rev. Lett.}\ }\textbf {\bibinfo {volume} {119}},\ \bibinfo
  {pages} {027001} (\bibinfo {year} {2017})}\BibitemShut {NoStop}%
\bibitem [{\citenamefont {Watanabe}\ and\ \citenamefont
  {Yanase}(2017)}]{Watanabe2017}%
  \BibitemOpen
  \bibfield  {author} {\bibinfo {author} {\bibfnamefont {H.}~\bibnamefont
  {Watanabe}}\ and\ \bibinfo {author} {\bibfnamefont {Y.}~\bibnamefont
  {Yanase}},\ }\href {\doibase 10.1103/PhysRevB.96.064432} {\bibfield
  {journal} {\bibinfo  {journal} {Phys. Rev. B}\ }\textbf {\bibinfo {volume}
  {96}},\ \bibinfo {pages} {064432} (\bibinfo {year} {2017})}\BibitemShut
  {NoStop}%
\bibitem [{\citenamefont {Fu}(2015)}]{Fu}%
  \BibitemOpen
  \bibfield  {author} {\bibinfo {author} {\bibfnamefont {L.}~\bibnamefont
  {Fu}},\ }\href {\doibase 10.1103/PhysRevLett.115.026401} {\bibfield
  {journal} {\bibinfo  {journal} {Phys. Rev. Lett.}\ }\textbf {\bibinfo
  {volume} {115}},\ \bibinfo {pages} {026401} (\bibinfo {year}
  {2015})}\BibitemShut {NoStop}%
\bibitem [{\citenamefont {Staub}\ \emph {et~al.}(2009)\citenamefont {Staub},
  \citenamefont {Bodenthin}, \citenamefont {Piamonteze}, \citenamefont
  {Garc\'{\i}a-Fern\'andez}, \citenamefont {Scagnoli}, \citenamefont
  {Garganourakis}, \citenamefont {Koohpayeh}, \citenamefont {Fort},\ and\
  \citenamefont {Lovesey}}]{Staub2009}%
  \BibitemOpen
  \bibfield  {author} {\bibinfo {author} {\bibfnamefont {U.}~\bibnamefont
  {Staub}}, \bibinfo {author} {\bibfnamefont {Y.}~\bibnamefont {Bodenthin}},
  \bibinfo {author} {\bibfnamefont {C.}~\bibnamefont {Piamonteze}}, \bibinfo
  {author} {\bibfnamefont {M.}~\bibnamefont {Garc\'{\i}a-Fern\'andez}},
  \bibinfo {author} {\bibfnamefont {V.}~\bibnamefont {Scagnoli}}, \bibinfo
  {author} {\bibfnamefont {M.}~\bibnamefont {Garganourakis}}, \bibinfo {author}
  {\bibfnamefont {S.}~\bibnamefont {Koohpayeh}}, \bibinfo {author}
  {\bibfnamefont {D.}~\bibnamefont {Fort}}, \ and\ \bibinfo {author}
  {\bibfnamefont {S.~W.}\ \bibnamefont {Lovesey}},\ }\href {\doibase
  10.1103/PhysRevB.80.140410} {\bibfield  {journal} {\bibinfo  {journal} {Phys.
  Rev. B}\ }\textbf {\bibinfo {volume} {80}},\ \bibinfo {pages} {140410}
  (\bibinfo {year} {2009})}\BibitemShut {NoStop}%
\bibitem [{\citenamefont {Kimura}\ \emph {et~al.}(2020)\citenamefont {Kimura},
  \citenamefont {Katsuyoshi}, \citenamefont {Sawada}, \citenamefont {Kimura},\
  and\ \citenamefont {Kimura}}]{Kimura}%
  \BibitemOpen
  \bibfield  {author} {\bibinfo {author} {\bibfnamefont {K.}~\bibnamefont
  {Kimura}}, \bibinfo {author} {\bibfnamefont {T.}~\bibnamefont {Katsuyoshi}},
  \bibinfo {author} {\bibfnamefont {Y.}~\bibnamefont {Sawada}}, \bibinfo
  {author} {\bibfnamefont {S.}~\bibnamefont {Kimura}}, \ and\ \bibinfo {author}
  {\bibfnamefont {T.}~\bibnamefont {Kimura}},\ }\href {\doibase
  10.1038/s43246-020-0040-3} {\bibfield  {journal} {\bibinfo  {journal}
  {Communications Materials}\ }\textbf {\bibinfo {volume} {1}},\ \bibinfo
  {pages} {39} (\bibinfo {year} {2020})}\BibitemShut {NoStop}%
\bibitem [{\citenamefont {Lovesey}(2014)}]{Lovesey}%
  \BibitemOpen
  \bibfield  {author} {\bibinfo {author} {\bibfnamefont {S.~W.}\ \bibnamefont
  {Lovesey}},\ }\href {\doibase 10.1088/0953-8984/26/35/356001} {\bibfield
  {journal} {\bibinfo  {journal} {Journal of Physics: Condensed Matter}\
  }\textbf {\bibinfo {volume} {26}},\ \bibinfo {pages} {356001} (\bibinfo
  {year} {2014})}\BibitemShut {NoStop}%
\bibitem [{\citenamefont {Lovesey}\ \emph {et~al.}(2019)\citenamefont
  {Lovesey}, \citenamefont {Chatterji}, \citenamefont {Stunault}, \citenamefont
  {Khalyavin},\ and\ \citenamefont {van~der Laan}}]{Lovesey2019}%
  \BibitemOpen
  \bibfield  {author} {\bibinfo {author} {\bibfnamefont {S.~W.}\ \bibnamefont
  {Lovesey}}, \bibinfo {author} {\bibfnamefont {T.}~\bibnamefont {Chatterji}},
  \bibinfo {author} {\bibfnamefont {A.}~\bibnamefont {Stunault}}, \bibinfo
  {author} {\bibfnamefont {D.~D.}\ \bibnamefont {Khalyavin}}, \ and\ \bibinfo
  {author} {\bibfnamefont {G.}~\bibnamefont {van~der Laan}},\ }\href {\doibase
  10.1103/PhysRevLett.122.047203} {\bibfield  {journal} {\bibinfo  {journal}
  {Phys. Rev. Lett.}\ }\textbf {\bibinfo {volume} {122}},\ \bibinfo {pages}
  {047203} (\bibinfo {year} {2019})}\BibitemShut {NoStop}%
\bibitem [{\citenamefont {Collins}\ \emph {et~al.}(2016)\citenamefont
  {Collins}, \citenamefont {Laundy}, \citenamefont {Connolley}, \citenamefont
  {van~der Laan}, \citenamefont {Fabrizi}, \citenamefont {Janssen},
  \citenamefont {Cooper}, \citenamefont {Ebert},\ and\ \citenamefont
  {Mankovsky}}]{Collins2016}%
  \BibitemOpen
  \bibfield  {author} {\bibinfo {author} {\bibfnamefont {S.~P.}\ \bibnamefont
  {Collins}}, \bibinfo {author} {\bibfnamefont {D.}~\bibnamefont {Laundy}},
  \bibinfo {author} {\bibfnamefont {T.}~\bibnamefont {Connolley}}, \bibinfo
  {author} {\bibfnamefont {G.}~\bibnamefont {van~der Laan}}, \bibinfo {author}
  {\bibfnamefont {F.}~\bibnamefont {Fabrizi}}, \bibinfo {author} {\bibfnamefont
  {O.}~\bibnamefont {Janssen}}, \bibinfo {author} {\bibfnamefont {M.~J.}\
  \bibnamefont {Cooper}}, \bibinfo {author} {\bibfnamefont {H.}~\bibnamefont
  {Ebert}}, \ and\ \bibinfo {author} {\bibfnamefont {S.}~\bibnamefont
  {Mankovsky}},\ }\href {\doibase 10.1107/S2053273316000863} {\bibfield
  {journal} {\bibinfo  {journal} {Acta Crystallographica Section A}\ }\textbf
  {\bibinfo {volume} {72}},\ \bibinfo {pages} {197} (\bibinfo {year}
  {2016})}\BibitemShut {NoStop}%
\bibitem [{\citenamefont {Collins}()}]{LNPO}%
  \BibitemOpen
  \bibfield  {author} {\bibinfo {author} {\bibfnamefont {S.~P.}\ \bibnamefont
  {Collins}},\ }\href@noop {} {\enquote {\bibinfo {title} {{Anti-symmetric
  Compton scattering in LiNiPO$_4$: Towards a direct probe of the
  magneto-electric multipole moment}},}\ }\bibinfo {howpublished} {(to be
  submitted) 2021}\BibitemShut {NoStop}%
\bibitem [{\citenamefont {Ederer}\ and\ \citenamefont
  {Spaldin}(2007)}]{Claude}%
  \BibitemOpen
  \bibfield  {author} {\bibinfo {author} {\bibfnamefont {C.}~\bibnamefont
  {Ederer}}\ and\ \bibinfo {author} {\bibfnamefont {N.~A.}\ \bibnamefont
  {Spaldin}},\ }\href {\doibase 10.1103/PhysRevB.76.214404} {\bibfield
  {journal} {\bibinfo  {journal} {Phys. Rev. B}\ }\textbf {\bibinfo {volume}
  {76}},\ \bibinfo {pages} {214404} (\bibinfo {year} {2007})}\BibitemShut
  {NoStop}%
\bibitem [{\citenamefont {Watanabe}\ and\ \citenamefont
  {Yanase}(2018)}]{Watanabe2018}%
  \BibitemOpen
  \bibfield  {author} {\bibinfo {author} {\bibfnamefont {H.}~\bibnamefont
  {Watanabe}}\ and\ \bibinfo {author} {\bibfnamefont {Y.}~\bibnamefont
  {Yanase}},\ }\href {\doibase 10.1103/PhysRevB.98.245129} {\bibfield
  {journal} {\bibinfo  {journal} {Phys. Rev. B}\ }\textbf {\bibinfo {volume}
  {98}},\ \bibinfo {pages} {245129} (\bibinfo {year} {2018})}\BibitemShut
  {NoStop}%
\bibitem [{\citenamefont {Mercier}\ and\ \citenamefont
  {Bauer}(1968)}]{Mercier}%
  \BibitemOpen
  \bibfield  {author} {\bibinfo {author} {\bibfnamefont {M.}~\bibnamefont
  {Mercier}}\ and\ \bibinfo {author} {\bibfnamefont {P.}~\bibnamefont
  {Bauer}},\ }\href@noop {} {\bibfield  {journal} {\bibinfo  {journal} {C. R.
  Acad. Sci. Paris}\ }\textbf {\bibinfo {volume} {267}},\ \bibinfo {pages}
  {465} (\bibinfo {year} {1968})}\BibitemShut {NoStop}%
\bibitem [{\citenamefont {Shick}\ \emph {et~al.}(2010)\citenamefont {Shick},
  \citenamefont {Khmelevskyi}, \citenamefont {Mryasov}, \citenamefont
  {Wunderlich},\ and\ \citenamefont {Jungwirth}}]{Shick}%
  \BibitemOpen
  \bibfield  {author} {\bibinfo {author} {\bibfnamefont {A.~B.}\ \bibnamefont
  {Shick}}, \bibinfo {author} {\bibfnamefont {S.}~\bibnamefont {Khmelevskyi}},
  \bibinfo {author} {\bibfnamefont {O.~N.}\ \bibnamefont {Mryasov}}, \bibinfo
  {author} {\bibfnamefont {J.}~\bibnamefont {Wunderlich}}, \ and\ \bibinfo
  {author} {\bibfnamefont {T.}~\bibnamefont {Jungwirth}},\ }\href {\doibase
  10.1103/PhysRevB.81.212409} {\bibfield  {journal} {\bibinfo  {journal} {Phys.
  Rev. B}\ }\textbf {\bibinfo {volume} {81}},\ \bibinfo {pages} {212409}
  (\bibinfo {year} {2010})}\BibitemShut {NoStop}%
\bibitem [{\citenamefont {Bodnar}\ \emph {et~al.}(2018)\citenamefont {Bodnar},
  \citenamefont {{\v S}mejkal}, \citenamefont {Turek}, \citenamefont
  {Jungwirth}, \citenamefont {Gomonay}, \citenamefont {Sinova}, \citenamefont
  {Sapozhnik}, \citenamefont {Elmers}, \citenamefont {Kl{\"a}ui},\ and\
  \citenamefont {Jourdan}}]{Bodnar}%
  \BibitemOpen
  \bibfield  {author} {\bibinfo {author} {\bibfnamefont {S.~Y.}\ \bibnamefont
  {Bodnar}}, \bibinfo {author} {\bibfnamefont {L.}~\bibnamefont {{\v
  S}mejkal}}, \bibinfo {author} {\bibfnamefont {I.}~\bibnamefont {Turek}},
  \bibinfo {author} {\bibfnamefont {T.}~\bibnamefont {Jungwirth}}, \bibinfo
  {author} {\bibfnamefont {O.}~\bibnamefont {Gomonay}}, \bibinfo {author}
  {\bibfnamefont {J.}~\bibnamefont {Sinova}}, \bibinfo {author} {\bibfnamefont
  {A.~A.}\ \bibnamefont {Sapozhnik}}, \bibinfo {author} {\bibfnamefont {H.~J.}\
  \bibnamefont {Elmers}}, \bibinfo {author} {\bibfnamefont {M.}~\bibnamefont
  {Kl{\"a}ui}}, \ and\ \bibinfo {author} {\bibfnamefont {M.}~\bibnamefont
  {Jourdan}},\ }\href {\doibase 10.1038/s41467-017-02780-x} {\bibfield
  {journal} {\bibinfo  {journal} {Nature Communications}\ }\textbf {\bibinfo
  {volume} {9}},\ \bibinfo {pages} {348} (\bibinfo {year} {2018})}\BibitemShut
  {NoStop}%
\bibitem [{\citenamefont {Chen}\ \emph {et~al.}(2019)\citenamefont {Chen},
  \citenamefont {Zhou}, \citenamefont {Cheng}, \citenamefont {Song},
  \citenamefont {Zhang}, \citenamefont {Wu}, \citenamefont {Ba}, \citenamefont
  {Li}, \citenamefont {Sun}, \citenamefont {You}, \citenamefont {Zhao},\ and\
  \citenamefont {Pan}}]{Chen}%
  \BibitemOpen
  \bibfield  {author} {\bibinfo {author} {\bibfnamefont {X.}~\bibnamefont
  {Chen}}, \bibinfo {author} {\bibfnamefont {X.}~\bibnamefont {Zhou}}, \bibinfo
  {author} {\bibfnamefont {R.}~\bibnamefont {Cheng}}, \bibinfo {author}
  {\bibfnamefont {C.}~\bibnamefont {Song}}, \bibinfo {author} {\bibfnamefont
  {J.}~\bibnamefont {Zhang}}, \bibinfo {author} {\bibfnamefont
  {Y.}~\bibnamefont {Wu}}, \bibinfo {author} {\bibfnamefont {Y.}~\bibnamefont
  {Ba}}, \bibinfo {author} {\bibfnamefont {H.}~\bibnamefont {Li}}, \bibinfo
  {author} {\bibfnamefont {Y.}~\bibnamefont {Sun}}, \bibinfo {author}
  {\bibfnamefont {Y.}~\bibnamefont {You}}, \bibinfo {author} {\bibfnamefont
  {Y.}~\bibnamefont {Zhao}}, \ and\ \bibinfo {author} {\bibfnamefont
  {F.}~\bibnamefont {Pan}},\ }\href {\doibase 10.1038/s41563-019-0424-2}
  {\bibfield  {journal} {\bibinfo  {journal} {Nature Materials}\ }\textbf
  {\bibinfo {volume} {18}},\ \bibinfo {pages} {931} (\bibinfo {year}
  {2019})}\BibitemShut {NoStop}%
\bibitem [{\citenamefont {Thöle}\ \emph {et~al.}(2020)\citenamefont {Thöle},
  \citenamefont {Keliri},\ and\ \citenamefont {Spaldin}}]{Florian}%
  \BibitemOpen
  \bibfield  {author} {\bibinfo {author} {\bibfnamefont {F.}~\bibnamefont
  {Thöle}}, \bibinfo {author} {\bibfnamefont {A.}~\bibnamefont {Keliri}}, \
  and\ \bibinfo {author} {\bibfnamefont {N.~A.}\ \bibnamefont {Spaldin}},\
  }\href {\doibase 10.1063/5.0006071} {\bibfield  {journal} {\bibinfo
  {journal} {Journal of Applied Physics}\ }\textbf {\bibinfo {volume} {127}},\
  \bibinfo {pages} {213905} (\bibinfo {year} {2020})},\ \Eprint
  {http://arxiv.org/abs/https://doi.org/10.1063/5.0006071}
  {https://doi.org/10.1063/5.0006071} \BibitemShut {NoStop}%
\bibitem [{\citenamefont {Abrahams}\ and\ \citenamefont
  {Easson}(1993)}]{Abrahams}%
  \BibitemOpen
  \bibfield  {author} {\bibinfo {author} {\bibfnamefont {I.}~\bibnamefont
  {Abrahams}}\ and\ \bibinfo {author} {\bibfnamefont {K.~S.}\ \bibnamefont
  {Easson}},\ }\href {\doibase 10.1107/S0108270192013064} {\bibfield  {journal}
  {\bibinfo  {journal} {Acta Crystallographica Section C}\ }\textbf {\bibinfo
  {volume} {49}},\ \bibinfo {pages} {925} (\bibinfo {year} {1993})}\BibitemShut
  {NoStop}%
\bibitem [{\citenamefont {Fogh}\ \emph {et~al.}(2020)\citenamefont {Fogh},
  \citenamefont {Kihara}, \citenamefont {Toft-Petersen}, \citenamefont
  {Bartkowiak}, \citenamefont {Narumi}, \citenamefont {Prokhnenko},
  \citenamefont {Miyake}, \citenamefont {Tokunaga}, \citenamefont {Oikawa},
  \citenamefont {S\o{}rensen}, \citenamefont {Dyrnum}, \citenamefont {Grimmer},
  \citenamefont {Nojiri},\ and\ \citenamefont {Christensen}}]{Exp2020}%
  \BibitemOpen
  \bibfield  {author} {\bibinfo {author} {\bibfnamefont {E.}~\bibnamefont
  {Fogh}}, \bibinfo {author} {\bibfnamefont {T.}~\bibnamefont {Kihara}},
  \bibinfo {author} {\bibfnamefont {R.}~\bibnamefont {Toft-Petersen}}, \bibinfo
  {author} {\bibfnamefont {M.}~\bibnamefont {Bartkowiak}}, \bibinfo {author}
  {\bibfnamefont {Y.}~\bibnamefont {Narumi}}, \bibinfo {author} {\bibfnamefont
  {O.}~\bibnamefont {Prokhnenko}}, \bibinfo {author} {\bibfnamefont
  {A.}~\bibnamefont {Miyake}}, \bibinfo {author} {\bibfnamefont
  {M.}~\bibnamefont {Tokunaga}}, \bibinfo {author} {\bibfnamefont
  {K.}~\bibnamefont {Oikawa}}, \bibinfo {author} {\bibfnamefont {M.~K.}\
  \bibnamefont {S\o{}rensen}}, \bibinfo {author} {\bibfnamefont {J.~C.}\
  \bibnamefont {Dyrnum}}, \bibinfo {author} {\bibfnamefont {H.}~\bibnamefont
  {Grimmer}}, \bibinfo {author} {\bibfnamefont {H.}~\bibnamefont {Nojiri}}, \
  and\ \bibinfo {author} {\bibfnamefont {N.~B.}\ \bibnamefont {Christensen}},\
  }\href {\doibase 10.1103/PhysRevB.101.024403} {\bibfield  {journal} {\bibinfo
   {journal} {Phys. Rev. B}\ }\textbf {\bibinfo {volume} {101}},\ \bibinfo
  {pages} {024403} (\bibinfo {year} {2020})}\BibitemShut {NoStop}%
\bibitem [{\citenamefont {Jensen}\ \emph {et~al.}(2009)\citenamefont {Jensen},
  \citenamefont {Christensen}, \citenamefont {Kenzelmann}, \citenamefont
  {R\o{}nnow}, \citenamefont {Niedermayer}, \citenamefont {Andersen},
  \citenamefont {Lefmann}, \citenamefont {Schefer}, \citenamefont
  {v.~Zimmermann}, \citenamefont {Li}, \citenamefont {Zarestky},\ and\
  \citenamefont {Vaknin}}]{Jensen}%
  \BibitemOpen
  \bibfield  {author} {\bibinfo {author} {\bibfnamefont {T.~B.~S.}\
  \bibnamefont {Jensen}}, \bibinfo {author} {\bibfnamefont {N.~B.}\
  \bibnamefont {Christensen}}, \bibinfo {author} {\bibfnamefont
  {M.}~\bibnamefont {Kenzelmann}}, \bibinfo {author} {\bibfnamefont {H.~M.}\
  \bibnamefont {R\o{}nnow}}, \bibinfo {author} {\bibfnamefont {C.}~\bibnamefont
  {Niedermayer}}, \bibinfo {author} {\bibfnamefont {N.~H.}\ \bibnamefont
  {Andersen}}, \bibinfo {author} {\bibfnamefont {K.}~\bibnamefont {Lefmann}},
  \bibinfo {author} {\bibfnamefont {J.}~\bibnamefont {Schefer}}, \bibinfo
  {author} {\bibfnamefont {M.}~\bibnamefont {v.~Zimmermann}}, \bibinfo {author}
  {\bibfnamefont {J.}~\bibnamefont {Li}}, \bibinfo {author} {\bibfnamefont
  {J.~L.}\ \bibnamefont {Zarestky}}, \ and\ \bibinfo {author} {\bibfnamefont
  {D.}~\bibnamefont {Vaknin}},\ }\href {\doibase 10.1103/PhysRevB.79.092412}
  {\bibfield  {journal} {\bibinfo  {journal} {Phys. Rev. B}\ }\textbf {\bibinfo
  {volume} {79}},\ \bibinfo {pages} {092412} (\bibinfo {year}
  {2009})}\BibitemShut {NoStop}%
\bibitem [{\citenamefont {Inui}\ \emph {et~al.}(1990)\citenamefont {Inui},
  \citenamefont {Tanabe},\ and\ \citenamefont {Onodera}}]{books}%
  \BibitemOpen
  \bibfield  {author} {\bibinfo {author} {\bibfnamefont {T.}~\bibnamefont
  {Inui}}, \bibinfo {author} {\bibfnamefont {Y.}~\bibnamefont {Tanabe}}, \ and\
  \bibinfo {author} {\bibfnamefont {Y.}~\bibnamefont {Onodera}},\ }\href@noop
  {} {\bibfield  {journal} {\bibinfo  {journal} {Springer, Berlin}\ }\textbf
  {\bibinfo {volume} {78}} (\bibinfo {year} {1990})}\BibitemShut {NoStop}%
\bibitem [{\citenamefont {Perez-Mato}\ \emph {et~al.}(2015)\citenamefont
  {Perez-Mato}, \citenamefont {Gallego}, \citenamefont {Tasci}, \citenamefont
  {Elcoro}, \citenamefont {de~la Flor},\ and\ \citenamefont {Aroyo}}]{Bilbao}%
  \BibitemOpen
  \bibfield  {author} {\bibinfo {author} {\bibfnamefont {J.}~\bibnamefont
  {Perez-Mato}}, \bibinfo {author} {\bibfnamefont {S.}~\bibnamefont {Gallego}},
  \bibinfo {author} {\bibfnamefont {E.}~\bibnamefont {Tasci}}, \bibinfo
  {author} {\bibfnamefont {L.}~\bibnamefont {Elcoro}}, \bibinfo {author}
  {\bibfnamefont {G.}~\bibnamefont {de~la Flor}}, \ and\ \bibinfo {author}
  {\bibfnamefont {M.}~\bibnamefont {Aroyo}},\ }\href {\doibase
  10.1146/annurev-matsci-070214-021008} {\bibfield  {journal} {\bibinfo
  {journal} {Annual Review of Materials Research}\ }\textbf {\bibinfo {volume}
  {45}},\ \bibinfo {pages} {217} (\bibinfo {year} {2015})},\ \Eprint
  {http://arxiv.org/abs/https://doi.org/10.1146/annurev-matsci-070214-021008}
  {https://doi.org/10.1146/annurev-matsci-070214-021008} \BibitemShut {NoStop}%
\bibitem [{cod()}]{code}%
  \BibitemOpen
  \href@noop {} {\enquote {\bibinfo {title} {{The Elk Code}},}\ }\bibinfo
  {howpublished} {\url{http://elk.sourceforge.net/}}\BibitemShut {NoStop}%
\bibitem [{\citenamefont {Ernsting}\ \emph {et~al.}(2014)\citenamefont
  {Ernsting}, \citenamefont {Billington}, \citenamefont {Haynes}, \citenamefont
  {Millichamp}, \citenamefont {Taylor}, \citenamefont {Duffy}, \citenamefont
  {Giblin}, \citenamefont {Dewhurst},\ and\ \citenamefont {Dugdale}}]{elk}%
  \BibitemOpen
  \bibfield  {author} {\bibinfo {author} {\bibfnamefont {D.}~\bibnamefont
  {Ernsting}}, \bibinfo {author} {\bibfnamefont {D.}~\bibnamefont
  {Billington}}, \bibinfo {author} {\bibfnamefont {T.~D.}\ \bibnamefont
  {Haynes}}, \bibinfo {author} {\bibfnamefont {T.~E.}\ \bibnamefont
  {Millichamp}}, \bibinfo {author} {\bibfnamefont {J.~W.}\ \bibnamefont
  {Taylor}}, \bibinfo {author} {\bibfnamefont {J.~A.}\ \bibnamefont {Duffy}},
  \bibinfo {author} {\bibfnamefont {S.~R.}\ \bibnamefont {Giblin}}, \bibinfo
  {author} {\bibfnamefont {J.~K.}\ \bibnamefont {Dewhurst}}, \ and\ \bibinfo
  {author} {\bibfnamefont {S.~B.}\ \bibnamefont {Dugdale}},\ }\href {\doibase
  10.1088/0953-8984/26/49/495501} {\bibfield  {journal} {\bibinfo  {journal}
  {Journal of Physics: Condensed Matter}\ }\textbf {\bibinfo {volume} {26}},\
  \bibinfo {pages} {495501} (\bibinfo {year} {2014})}\BibitemShut {NoStop}%
\bibitem [{\citenamefont {Biggs}\ \emph {et~al.}(1975)\citenamefont {Biggs},
  \citenamefont {Mendelsohn},\ and\ \citenamefont {Mann}}]{Biggs}%
  \BibitemOpen
  \bibfield  {author} {\bibinfo {author} {\bibfnamefont {F.}~\bibnamefont
  {Biggs}}, \bibinfo {author} {\bibfnamefont {L.}~\bibnamefont {Mendelsohn}}, \
  and\ \bibinfo {author} {\bibfnamefont {J.}~\bibnamefont {Mann}},\ }\href
  {\doibase https://doi.org/10.1016/0092-640X(75)90030-3} {\bibfield  {journal}
  {\bibinfo  {journal} {Atomic Data and Nuclear Data Tables}\ }\textbf
  {\bibinfo {volume} {16}},\ \bibinfo {pages} {201} (\bibinfo {year}
  {1975})}\BibitemShut {NoStop}%
\bibitem [{\citenamefont {Wells}\ and\ \citenamefont {Smith}(1970)}]{Wells}%
  \BibitemOpen
  \bibfield  {author} {\bibinfo {author} {\bibfnamefont {P.}~\bibnamefont
  {Wells}}\ and\ \bibinfo {author} {\bibfnamefont {J.~H.}\ \bibnamefont
  {Smith}},\ }\href {\doibase https://doi.org/10.1107/S056773947000092X}
  {\bibfield  {journal} {\bibinfo  {journal} {Acta Crystallographica Section
  A}\ }\textbf {\bibinfo {volume} {26}},\ \bibinfo {pages} {379} (\bibinfo
  {year} {1970})}\BibitemShut {NoStop}%
\bibitem [{\citenamefont {Elmers}\ \emph {et~al.}(2020)\citenamefont {Elmers},
  \citenamefont {Chernov}, \citenamefont {D'Souza}, \citenamefont
  {Bommanaboyena}, \citenamefont {Bodnar}, \citenamefont {Medjanik},
  \citenamefont {Babenkov}, \citenamefont {Fedchenko}, \citenamefont
  {Vasilyev}, \citenamefont {Agustsson}, \citenamefont {Schlueter},
  \citenamefont {Gloskovskii}, \citenamefont {Matveyev}, \citenamefont
  {Strocov}, \citenamefont {Skourski}, \citenamefont {{\v S}mejkal},
  \citenamefont {Sinova}, \citenamefont {Min{\'a}r}, \citenamefont {Kl{\"a}ui},
  \citenamefont {Sch{\"o}nhense},\ and\ \citenamefont {Jourdan}}]{Elmers}%
  \BibitemOpen
  \bibfield  {author} {\bibinfo {author} {\bibfnamefont {H.~J.}\ \bibnamefont
  {Elmers}}, \bibinfo {author} {\bibfnamefont {S.~V.}\ \bibnamefont {Chernov}},
  \bibinfo {author} {\bibfnamefont {S.~W.}\ \bibnamefont {D'Souza}}, \bibinfo
  {author} {\bibfnamefont {S.~P.}\ \bibnamefont {Bommanaboyena}}, \bibinfo
  {author} {\bibfnamefont {S.~Y.}\ \bibnamefont {Bodnar}}, \bibinfo {author}
  {\bibfnamefont {K.}~\bibnamefont {Medjanik}}, \bibinfo {author}
  {\bibfnamefont {S.}~\bibnamefont {Babenkov}}, \bibinfo {author}
  {\bibfnamefont {O.}~\bibnamefont {Fedchenko}}, \bibinfo {author}
  {\bibfnamefont {D.}~\bibnamefont {Vasilyev}}, \bibinfo {author}
  {\bibfnamefont {S.~Y.}\ \bibnamefont {Agustsson}}, \bibinfo {author}
  {\bibfnamefont {C.}~\bibnamefont {Schlueter}}, \bibinfo {author}
  {\bibfnamefont {A.}~\bibnamefont {Gloskovskii}}, \bibinfo {author}
  {\bibfnamefont {Y.}~\bibnamefont {Matveyev}}, \bibinfo {author}
  {\bibfnamefont {V.~N.}\ \bibnamefont {Strocov}}, \bibinfo {author}
  {\bibfnamefont {Y.}~\bibnamefont {Skourski}}, \bibinfo {author}
  {\bibfnamefont {L.}~\bibnamefont {{\v S}mejkal}}, \bibinfo {author}
  {\bibfnamefont {J.}~\bibnamefont {Sinova}}, \bibinfo {author} {\bibfnamefont
  {J.}~\bibnamefont {Min{\'a}r}}, \bibinfo {author} {\bibfnamefont
  {M.}~\bibnamefont {Kl{\"a}ui}}, \bibinfo {author} {\bibfnamefont
  {G.}~\bibnamefont {Sch{\"o}nhense}}, \ and\ \bibinfo {author} {\bibfnamefont
  {M.}~\bibnamefont {Jourdan}},\ }\bibfield  {booktitle} {\emph {\bibinfo
  {booktitle} {ACS Nano}},\ }\href {\doibase 10.1021/acsnano.0c08215}
  {\bibfield  {journal} {\bibinfo  {journal} {ACS Nano}\ }\textbf {\bibinfo
  {volume} {14}},\ \bibinfo {pages} {17554} (\bibinfo {year}
  {2020})}\BibitemShut {NoStop}%
\bibitem [{\citenamefont {Sapozhnik}\ \emph {et~al.}(2018)\citenamefont
  {Sapozhnik}, \citenamefont {Filianina}, \citenamefont {Bodnar}, \citenamefont
  {Lamirand}, \citenamefont {Mawass}, \citenamefont {Skourski}, \citenamefont
  {Elmers}, \citenamefont {Zabel}, \citenamefont {Kl\"aui},\ and\ \citenamefont
  {Jourdan}}]{Sapozhnik}%
  \BibitemOpen
  \bibfield  {author} {\bibinfo {author} {\bibfnamefont {A.~A.}\ \bibnamefont
  {Sapozhnik}}, \bibinfo {author} {\bibfnamefont {M.}~\bibnamefont
  {Filianina}}, \bibinfo {author} {\bibfnamefont {S.~Y.}\ \bibnamefont
  {Bodnar}}, \bibinfo {author} {\bibfnamefont {A.}~\bibnamefont {Lamirand}},
  \bibinfo {author} {\bibfnamefont {M.-A.}\ \bibnamefont {Mawass}}, \bibinfo
  {author} {\bibfnamefont {Y.}~\bibnamefont {Skourski}}, \bibinfo {author}
  {\bibfnamefont {H.-J.}\ \bibnamefont {Elmers}}, \bibinfo {author}
  {\bibfnamefont {H.}~\bibnamefont {Zabel}}, \bibinfo {author} {\bibfnamefont
  {M.}~\bibnamefont {Kl\"aui}}, \ and\ \bibinfo {author} {\bibfnamefont
  {M.}~\bibnamefont {Jourdan}},\ }\href {\doibase 10.1103/PhysRevB.97.134429}
  {\bibfield  {journal} {\bibinfo  {journal} {Phys. Rev. B}\ }\textbf {\bibinfo
  {volume} {97}},\ \bibinfo {pages} {134429} (\bibinfo {year}
  {2018})}\BibitemShut {NoStop}%
\end{thebibliography}%

\end{document}